\documentclass[%
superscriptaddress,
amsmath,amssymb,
aps,pra,
twocolumn,
floatfix,
]{revtex4}

\usepackage{graphicx}% Include figure files
\usepackage{dcolumn}% Align table columns on decimal point
\usepackage{bm}% bold math
\usepackage{hyperref}% add hypertext capabilities
\usepackage{amsmath,amsfonts,amssymb,amsthm}
\usepackage{braket}
\usepackage{slashed}
\usepackage{comment}
\usepackage{float}
\usepackage{mhchem}
\usepackage{tabularx}
\def \be {\begin{equation}}
\def \ee {\end{equation}}
\usepackage{times}
\usepackage{nicefrac}
\usepackage{epsf}
\usepackage{bm}
\usepackage{bbm}
\usepackage{color}

\preprint{APS/123-QED}
\begin{document}
\title{Redefined vacuum approach and gauge-invariant subsets in two-photon-exchange diagrams for a closed-shell system with a valence electron}
% Force line breaks with \\
%\thanks{A footnote to the article title}%
%}
\author{R. N. Soguel}
\email{romain.soguel@uni-jena.de}
\affiliation{Theoretisch-Physikalisches Institut, Friedrich-Schiller-Universität Jena, Max-Wien-Platz 1, 07743 Jena, Germany}
\affiliation{Helmholtz-Institut Jena, Fr\"obelstieg 3, 07743 Jena, Germany}
\affiliation{GSI Helmholtzzentrum für Schwerionenforschung GmbH, Planckstraße 1, 64291 Darmstadt, Germany}
\author{A. V. Volotka}
\affiliation{Department of Physics and Engineering, ITMO University, Kronverskiy pr. 49, 197101 St. Petersburg, Russia}
\affiliation{Helmholtz-Institut Jena, Fr\"obelstieg 3, 07743 Jena, Germany}
\affiliation{GSI Helmholtzzentrum für Schwerionenforschung GmbH, Planckstraße 1, 64291 Darmstadt, Germany}
\author{E. V. Tryapitsyna}
\affiliation{Department of Physics, St. Petersburg State University, Universitetskaya nab. 7/9, 199034 St. Petersburg, Russia}
\author{D. A. Glazov}
\affiliation{Department of Physics, St. Petersburg State University, Universitetskaya nab. 7/9, 199034 St. Petersburg, Russia}
\author{V. P. Kosheleva}
\affiliation{Theoretisch-Physikalisches Institut, Friedrich-Schiller-Universität Jena, Max-Wien-Platz 1, 07743 Jena, Germany}
\affiliation{Helmholtz-Institut Jena, Fr\"obelstieg 3, 07743 Jena, Germany}
\affiliation{GSI Helmholtzzentrum für Schwerionenforschung GmbH, Planckstraße 1, 64291 Darmstadt, Germany}
\author{S. Fritzsche}
\affiliation{Theoretisch-Physikalisches Institut, Friedrich-Schiller-Universität Jena, Max-Wien-Platz 1, 07743 Jena, Germany}
\affiliation{Helmholtz-Institut Jena, Fr\"obelstieg 3, 07743 Jena, Germany}
\affiliation{GSI Helmholtzzentrum für Schwerionenforschung GmbH, Planckstraße 1, 64291 Darmstadt, Germany}
\date{\today}

\begin{abstract}
The two-photon-exchange diagrams for atoms with single valence electron are investigated. Calculation formulas are derived for an arbitrary state within the rigorous bound-state QED framework utilizing the redefined vacuum formalism. In contrast to other methods, the redefined vacuum approach enables the identification of eight gauge-invariant subsets and, thus, efficiently check the consistency of the obtained results. The gauge invariance of found subsets is demonstrated both analytically (for an arbitrary state) as well as numerically for $2s$, $2p_{1/2}$, and $2p_{3/2}$ valence electron in Li-like ions. Identifying gauge-invariant subsets in the framework of the proposed approach opens a way to tackle more complex diagrams, e.g., three-photon exchange, where the fragmentation on simpler subsets is crucial for its successful calculation.
\end{abstract}
\maketitle 

%=================================================================
%
\section{\label{sec:level1} Introduction}
The treatment of the interelectronic interaction remains a cornerstone for accurate theoretical predictions of the energy levels in many-electron atoms or ions. The relativistic many-body perturbation theory (MBPT) or configuration-interaction calculations usually treat the electron-electron interaction within the Breit approximation on the basis of the so-called \emph{no-pair} Hamiltonian \cite{faustov:1970:478, sucher:1980:348}. However, the use of this approximation allows one to evaluate energy levels accurately only up to the order $(\alpha Z)^2$ in atomic units. Here, $\alpha$ is the fine structure constant and $Z$ is the nuclear charge or \emph{effective} charge in the case of the outer electrons. In order to account for higher-order effects, one has to employ the bound-state quantum electrodynamics (QED) formalism. The QED treatment of the interelectronic interaction yields additional corrections of the order $(\alpha Z)^3$ and higher, which in many cases become comparable with experimental uncertainty and the accuracy reached for correlation effects.

The straightforward examples are highly charged ions, where $Z$ is large and the QED corrections mentioned above enter on the same footing as the correlation effects. As a consequence, the $(\alpha Z)^3$ corrections beyond the no-pair Hamiltonian should be taken into account by theory in order to achieve an agreement with the high-precision experiments for He-like \cite{gumberidze:2004:203004, bruhns:2007:113001, amaro:2012:043005}, Li-like \cite{brandau:2003:073202, beiersdorfer:2005:233003}, or B-like \cite{draganic:2003:183001, beiersdorfer:1998:1944} ions. Other examples are many-electron ions where the correlation energies can be grasped rather accurately, such as F-like \cite{li:2018:R020502, volotka:2019:010502}, Na-like \cite{johnson:1988:2699}, Mg-like \cite{chen:1997:3440}, Al-like \cite{safronova:2002:022507}, Co-like \cite{si:2018:012504}, and other ions.  

Within the bound-state QED, the interelectronic interaction is usually treated perturbatively as an expansion over the number of exchanged photons. While the first-order contribution, which corresponds to the one-photon exchange, is relatively simple to deal with, most of the attention is paid to the next order, namely, the two-photon exchange. The two-photon exchange diagrams were first calculated in the milestone paper by Blundell \textit{et al.} \cite{blundell:1993:2615} for the ground state of He-like ions. Later, the results were confirmed by several groups and the calculations were extended to their excited states \cite{lindgren:1995:1167, mohr:2000:052501, andreev:2001:042513, andreev:2003:012503, artemyev:2005:062104, malyshev:2020:010501}. Presently, the two-photon exchange has been also evaluated for Li-like \cite{yerokhin:2001:032109, andreev:2001:042513, andreev:2003:012503, artemyev:2003:062506, yerokhin:2007:062501, kozhedub:2010:042513, sapirstein:2011:012504}, Be-like \cite{malyshev:2014:062517, malyshev:2015:012514}, B-like \cite{artemyev:2007:173004, artemyev:2013:032518, malyshev:2017:022512}, and Na-like \cite{sapirstein:2015:062508} ions. The current experimental and theoretical developments recently reviewed by Indelicato \cite{indelicato:2019:232001} suggest the necessity to extend these computations also to other systems with more complicated electronic structures. An essential step towards this goal is to derive the computational formulas for the corresponding diagrams. All the derivations performed so far used a zeroth-order many-electron wave-function constructed as a Slater determinant (or sum of Slater determinants) with all electrons involved \cite{blundell:1993:2615, shabaev:1994:4489, sapirstein:2015:062508}. Such a derivation becomes increasingly difficult for many-electron systems. The vacuum redefinition in QED, which is extensively used in MBPT to describe the states with many electrons involved, could be a path towards an extension of two-photon-exchange calculations to other ions and atoms.

The current study reports a derivation of the calculation formulas for the two-photon-exchange diagrams for atoms with one electron over closed shells, based on the vacuum redefinition formalism. The one-electron two-loop diagrams serve as a starting point for this derivation. The advantages of the method are as follows. First, the one-electron frame makes formulas valid for any system with one valence electron configuration. Second, translation of the two-loop gauge-invariant subsets into the generic many-electron system allows us to identify simple gauge-invariant subsets. Based on those subsets, one can split the two-photon-exchange contributions into two different two-electron parts, direct and exchange, and six other three-electron parts plus counterpotential terms in the case of an extended Furry picture. Each of the identified subsets is gauge invariant due to the corresponding two-loop one-electron origin. The gauge invariance is also demonstrated analytically for a general state and numerically for the $2s$, $2p_{1/2}$, and $2p_{3/2}$ valence electron configurations in Li-like ions. In this respect, the effectiveness of the vacuum redefinition method is shown for a closed-shell system with a valence electron.

The paper is organized as follows. In Section II, a description of the computational techniques and methodology used is provided. Section III illustrates the method with the one-photon-exchange diagram.  In Section IV, a partition among the different two-loop diagrams is presented. The formulas for the two-photon exchange are derived and the gauge invariance of the three-electron part is demonstrated analytically for each subset. In Section V, the total energy correction due to the two-photon exchange is presented, and numerical evaluation for each gauge-invariant subset in Feynman and Coulomb gauges is provided. Finally, conclusions are given in Section VI. 
\\Natural units ($\hbar = c = m_e=1$) are used throughout this paper; the fine-structure constant is defined as $\alpha = e^{2}/(4\pi), e<0$.  The metric tensor is taken to be $\eta^{\mu \nu}= \eta_{\mu \nu}= \text{diag}(1,-1,-1,-1)$. Unless explicitly stated, all integrals are meant to be on the interval  $\ ] -\infty, \infty \ [$.

%=================================================================
%
\section{Method and formulation}
\subsection{QED in Furry picture}
The relativistic theory of the electron is based on the Dirac equation,
\begin{equation} 
h_{\text{D}} \phi_j(\boldsymbol{x}) = [-i \boldsymbol{\alpha} \cdot \boldsymbol{\nabla} + \beta  + V(\boldsymbol{x})]\phi_j(\boldsymbol{x})= \epsilon_j \phi_j(\boldsymbol{x})\,,
\label{Dirac_eq}
\end{equation}
where $\alpha^{k}$, $\beta$ are Dirac matrices and $j$ uniquely characterizes the solution, i.e., stands for all quantum numbers. $\phi_j$ is the static solution, which yields the corresponding time-dependent solution when being multiplied by the $e^{-i\epsilon_j x^0}$ phase. Within the Furry picture \cite{furry:1951:115}, eigenstates are considered to be solutions of the Dirac equation in the presence of an external classical field. The original Furry picture considers the Coulomb potential $ V_C(\boldsymbol{x})$ created by a nucleus, $V(\boldsymbol{x}) = V_C(\boldsymbol{x})$. The extended Furry picture encapsulates a screening potential $U(\boldsymbol{x}) $ in addition to the Coulomb one, i.e., $V(\boldsymbol{x}) = V_C(\boldsymbol{x}) + U(\boldsymbol{x})$. The unperturbed normal ordered Hamiltonian is given by \cite{mohr:1998:227}
\begin{equation}
H_0 = \int d^{3}x : \psi^{(0)\dagger}(x) h_{\text{D}} \psi^{(0)} (x):\,,
\end{equation}
where the fermion field operator is expanded in terms of creation and annihilation operators
\begin{equation}
 \psi^{(0)}(x) = \sum_{\epsilon_j > E^F } a_j \phi_j(\boldsymbol{x}) e^{-i\epsilon_j x^0} + \sum_{\epsilon_j < E^F } b_j^{\dagger } \phi_j(\boldsymbol{x}) e^{-i\epsilon_j x^0}\,.
\label{bound_exp}
\end{equation}
The Fermi level $E^F$ is set to $E^F = 0$ separating the Dirac sea from the rest of the spectrum. The valence electron state is described in the Fock space as 
\begin{equation}
\ket{v a b ...}= a_v^{\dagger} a_{a}^{\dagger}  a_{b}^{\dagger}...\ket{0}\,.
\end{equation}
Here and in what follows the MBPT notations of Lindgren and Morisson \cite{lindgren_morrison} and Johnson \cite{johnson_2007} will be used: $v$ designates the valence electron, $a,b,...$ stand for core orbitals, and $i,j,...$ correspond to arbitrary states. The zero-order energy is given by
\begin{equation}
E_{v a b ...}^{(0)}= \bra{v a b ...} H_0\ket{v a b ...} = \epsilon_v + \epsilon_a + \epsilon_b +... \,.
\end{equation}
\subsection{Vacuum redefinition}
When the interest stays in the valence electron binding energy $ E^{(0)}_v$ or in its excitation energy, each being, respectively, 
\begin{eqnarray}
&&E^{(0)}_v = E_{v a b ...}^{(0)} - E^{(0)}_{ a b ...}\,,\nonumber\\
&&E^{(0)}_{v_1 a b ...} - E^{(0)}_{ v_2 a b ...} = E^{(0)}_{v_1} - E^{(0)}_{v_2}\,,
\end{eqnarray}
it is convenient to employ the concept of vacuum redefinition. A new vacuum state is chosen such that all core orbitals are occupied and the remaining ones are free \cite{lindgren_morrison}. Let us denote it by $\ket{\alpha}$ notation,
\begin{equation}
\ket{\alpha} = a_{a}^{\dagger} a_{b}^{\dagger} ... \ket{0}\,.  
\label{redefined vacuum}
\end{equation}
The meaning of creation and annihilation operators is changed for the core shell electrons ($a_a \rightarrow b_a^{\dagger}, \hspace{5pt} a_a^{\dagger} \rightarrow b_a$):
\begin{equation}
b_{a} \ket{\alpha} = 0\,,
\label{core-operators}
\end{equation}
and the fermion field operator now reads
\begin{equation}
\psi^{(0)}_{\alpha}(x)=  \sum_{\epsilon_j > E_{\alpha}^{F} } a_j \phi_j(\boldsymbol{x}) e^{-i\epsilon_j x^0} + \sum_{\epsilon_j < E_{\alpha}^{F} } b_j^{\dagger } \phi_j(\boldsymbol{x}) e^{-i\epsilon_j x^0}\,. 
\label{alpha bound state exp}
\end{equation}
The Fermi level now lies between the highest core state and the valence state, $E_{\alpha}^F \in (\epsilon_a, \epsilon_v)$. With this redefinition of the vacuum, the valence state is now described by
\begin{equation}
\ket{v} = a_v^{\dagger} \ket{\alpha}\,,
%\label{valences state}
\end{equation}
with zero-order binding energy
\begin{equation}
 E^{(0)}_v = \bra{v} H_0 \ket{v} = \epsilon_v\,.
\label{ezero}
\end{equation}
\subsection{Electron propagator}
The vacuum expectation value of the time-ordered product of two electron-positron field operators defines the electron propagator. For the standard vacuum using the bound state expansion in term of creation and annihilation operators (\ref{bound_exp}) one gets 
\begin{eqnarray}
&&\bra{0}  T\left[\psi^{(0)}(x) \bar{\psi}^{(0)}(y) \right] \ket{0}  \nonumber\\
&&=\frac{i}{2\pi} \int d\omega \sum_{j} \frac{\phi_j(\boldsymbol{x}) \bar{\phi}_j(\boldsymbol{y}) e^{-i(x^0 -y^0) \omega}}{\omega - \epsilon_j (1-i\varepsilon)}\,,
\end{eqnarray}
where $\varepsilon>0$ implies the limit to zero. For later use, let us define $u = 1 - i\varepsilon$. For the redefined vacuum the propagator can be written in the following form,
\begin{eqnarray}
&&\bra{\alpha}  T \left[\psi_{\alpha}^{(0)}(x) \bar{\psi}_{\alpha}^{(0)}(y) \right]\ket{\alpha}  
\nonumber\\
&& = \frac{i}{2\pi} \int d\omega \sum_{j} \frac{\phi_j(\boldsymbol{x}) \bar{\phi}_j(\boldsymbol{y}) e^{-i(x^0 -y^0) \omega}}{\omega - \epsilon_j + i\varepsilon (\epsilon_j-E_{\alpha}^{F}) }\,, 
\end{eqnarray}
which is also suitable for the standard vacuum, with the replacement $E_{\alpha}^{F} \to E^F=0$. 
% For latter purpose, let us define $u = 1 - i\varepsilon$. In the redefined vacuum case (\ref{alpha bound state exp}), we arrive to
% %
% \begin{eqnarray}
% &&\bra{\alpha}  T(\psi_{\alpha}^{(0)}(x) \bar{\psi}_{\alpha}^{(0)}(y))\ket{\alpha}  
% \nonumber\\
% && = \frac{i}{2\pi} \int d\omega \sum_{i} \frac{\phi_i(\boldsymbol{x}) \bar{\phi}_i(\boldsymbol{y}) e^{-i(x^0 -y^0) \omega}}{\omega - \epsilon_i u_{\alpha}}\,, \end{eqnarray}
% %
% with
% %
% \begin{eqnarray}
%  u_{\alpha}=  \left\{ \begin{matrix}   1 - i\varepsilon \equiv u, \hspace{10pt} &\epsilon_i \neq  \epsilon_a \\   1+ i \varepsilon, \hspace{10pt} &\epsilon_i =  \epsilon_a \end{matrix}\right.\,.    
% \end{eqnarray}
% %
Redefinition of the vacuum changes the poles circumvention prescription for the core states, so that they are relegated to the Dirac sea. It corresponds to a different integration path in the complex $\omega$ plane, see also Ref.~\cite{shabaev:2002:119}. In the following, we have to calculate the difference between these two propagators, it corresponds to a cut of the electron line on the diagram. Such a difference can be easily simplified by applying the Sokhotski-Plemelj theorem. The following equality is meant to be understood while integrating in the complex $\omega$ plane. For $p = 1, 2, \dots$ we have,
% %
% \begin{eqnarray}
% &&\sum_i\frac{\phi_i(\boldsymbol{x}) \bar{\phi}_i(\boldsymbol{x})}{(E - \omega - \epsilon_i u_{\alpha})^p} - \sum_i\frac{\phi_i(\boldsymbol{x}) \bar{\phi}_i(\boldsymbol{x})}{(E - \omega - \epsilon_i u)^p}    = \frac{2\pi i}{(p-1)!}\nonumber \\&& \times\frac{d^{(p-1)}}{d \omega^{(p-1)}}\sum_{a} \delta(E- \omega - \epsilon_{a})\phi_{a}(\boldsymbol{x}) \bar{\phi}_{a}(\boldsymbol{x})\,.
% \label{Sokhotski}
% \end{eqnarray}
% %
%
\begin{eqnarray}
&&\sum_j\frac{\phi_j(\boldsymbol{x}) \bar{\phi}_j(\boldsymbol{y})}{[\omega - \epsilon_j + i\varepsilon (\epsilon_j-E_{\alpha}^{F})]^p}
\nonumber \\&& 
-\sum_j\frac{\phi_j(\boldsymbol{x}) \bar{\phi}_j(\boldsymbol{y})}{[\omega - \epsilon_j + i\varepsilon (\epsilon_j-E^F)]^p}
\nonumber \\&& 
= \frac{2\pi i (-1)^p}{(p-1)!}\frac{d^{(p-1)}}{d \omega^{(p-1)}}\sum_{a} \delta(\omega - \epsilon_{a})\phi_{a}(\boldsymbol{x}) \bar{\phi}_{a}(\boldsymbol{y})\,.
\label{Sokhotski}
\end{eqnarray}
\subsection{Photon propagator}
Let us first introduce the following notations for the interelectronic-interaction operator $I$:
\begin{eqnarray}
&&I(\boldsymbol{x}- \boldsymbol{y}; \omega) = e^2 \alpha^{\mu} \alpha^{\nu} D_{\mu \nu} (\boldsymbol{x}- \boldsymbol{y}; \omega)\,, \nonumber\\ &&I^{\prime} (\boldsymbol{x}- \boldsymbol{y}; \omega) \equiv \frac{d I(\boldsymbol{x}- \boldsymbol{y}; \omega)}{d\omega}\,,
\end{eqnarray}
where $D_{\mu \nu} (\boldsymbol{x}- \boldsymbol{y};\omega)$ is the photon propagator and  $\alpha^{\mu} = (1, \boldsymbol{\alpha})$. In the Feynman and Coulomb gauges considered below, these operators fulfill the symmetry properties
\begin{eqnarray}
I(\boldsymbol{x}-\boldsymbol{y}; \omega) &=& I(\boldsymbol{x}-\boldsymbol{y};-\omega)\,,\nonumber\\
I^{\prime}(\boldsymbol{x}-\boldsymbol{y};\omega) &=& - I^{\prime}(\boldsymbol{x}-\boldsymbol{y};-\omega)\,,\nonumber\\
I^{\prime}(\boldsymbol{x}-\boldsymbol{y};0) &=& 0\,.
\end{eqnarray}
The following shorthand notation for the matrix element will be used throughout the whole paper:
\begin{equation}
I_{i j k l}(\omega) = \int d^{3}x d^{3}y \phi_i^{\dagger}(\boldsymbol{x}) \phi_j^{\dagger}(\boldsymbol{y}) I(\boldsymbol{x}- \boldsymbol{y};\omega)  \phi_k(\boldsymbol{x}) \phi_l(\boldsymbol{y})\,,
\end{equation}
for which the symmetry property holds
\begin{equation}
I_{i j k l }(\omega) = I_{ j i l k}(\omega)\,.
\label{eq:symmetry}
\end{equation}
%
% Furthermore, we define $\Delta_{ij} = \epsilon_i - \epsilon_j$.
%\subsubsection{Feynman gauge}
%
In the Feynman gauge, the photon propagator has the form, 
\begin{equation} 
D^{F}_{\mu\nu}(\boldsymbol{x}- \boldsymbol{y};\omega) =  \frac{\eta_{\mu \nu}}{4\pi|\boldsymbol{x}- \boldsymbol{y}|} e^{i \sqrt{\omega^2 + i\varepsilon} |\boldsymbol{x}- \boldsymbol{y}|}
\end{equation}
and the operator $I^{F}$ reads
\begin{eqnarray} 
I^{F}(\boldsymbol{x}- \boldsymbol{y};\omega) = \alpha \frac{1 - \boldsymbol{\alpha}_{x} \cdot \boldsymbol{\alpha}_{y} }{|\boldsymbol{x}- \boldsymbol{y}|} e^{i \tilde{\omega} |\boldsymbol{x}- \boldsymbol{y}|}\,,
\end{eqnarray}
with $\tilde{\omega} = \sqrt{\omega^2 + i \varepsilon}$ and the branch of the square root is fixed by the condition Im$(\tilde{\omega})>0$. Matrices $\boldsymbol{\alpha}_{x}$ and $\boldsymbol{\alpha}_{y}$ act on the Dirac bispinors with the arguments $\boldsymbol{x}$ and $\boldsymbol{y}$, respectively.
%
%\subsubsection{Coulomb gauge}
The photon propagator is given, in the Coulomb gauge, by
\begin{eqnarray}
D^{C}_{00}(\boldsymbol{x}-\boldsymbol{y};\omega) &=& \frac{1}{4\pi|\boldsymbol{x}-\boldsymbol{y}|}\,,\nonumber\\
D^{C}_{i0}(\boldsymbol{x}-\boldsymbol{y};\omega) &=& D^{C}_{0j}(\boldsymbol{x}-\boldsymbol{y};\omega)=0\,,\nonumber\\
D^{C}_{ij}(\boldsymbol{x}-\boldsymbol{y};\omega) &=& -\frac{\delta_{ij} e^{i\tilde{\omega}|\boldsymbol{x}-\boldsymbol{y}|}}{4\pi |\boldsymbol{x}-\boldsymbol{y}|} \nonumber\\ &-& \nabla_i^{(x)} \nabla_j^{(y)} \frac{1-e^{i\tilde{\omega} |\boldsymbol{x}-\boldsymbol{y}|}}{4\pi\omega^2|\boldsymbol{x}-\boldsymbol{y}|}\,. 
\end{eqnarray}
Thus, $I^C$ can be expressed as
\begin{align} 
I^C(\boldsymbol{x}- \boldsymbol{y};\omega) &= \alpha \left\{ \frac{1 - \boldsymbol{\alpha}_{x} \cdot \boldsymbol{\alpha}_{y} e^{i\tilde{\omega}|\boldsymbol{x}- \boldsymbol{y}|} }{|\boldsymbol{x}- \boldsymbol{y}|}  \right. \nonumber \\  &+ \left.  \left[ (\boldsymbol{\alpha}_{x} \cdot \boldsymbol{\nabla}_{x}), \left[ (\boldsymbol{\alpha}_{y} \cdot \boldsymbol{\nabla}_{y}), \frac{e^{i \tilde{\omega}|\boldsymbol{x}- \boldsymbol{y}|}-1} {\omega^2|\boldsymbol{x}- \boldsymbol{y}|} \right]\right]\right\}. \nonumber \\ 
\end{align}
%
%\subsubsection{Gauge difference}
%
In the following, we will demonstrate the gauge invariance of derived expressions. For this purpose, we are interested in the difference between the matrix elements with the photon propagator in the Feynman and Coulomb gauges. This difference is given by
\begin{eqnarray}
 \Delta I_{i j k l} (\omega) &\equiv& I_{i j k l}^{F}(\omega) -I_{i j k l}^{C}(\omega)     \nonumber \\ &=& ( \omega^2 + \Delta_{ik} \Delta_{jl} ) \tilde{I}_{i j k l}(\omega)\,,
\label{propagator_difference}
\end{eqnarray}
with
\begin{equation}
\tilde{I}(\boldsymbol{x}- \boldsymbol{y};\omega)=   \alpha \frac{e^{i \tilde{\omega} |\boldsymbol{x}- \boldsymbol{y}|} -1}{\omega^2|\boldsymbol{x}- \boldsymbol{y}|}
% \,. 
\end{equation}
and $\Delta_{ij} = \epsilon_i - \epsilon_j$.
We also need its derivative with respect to $\omega$, which can be cast as
\begin{equation} 
\Delta I^{\prime}_{i j k l} (\omega)= 2\omega \tilde{I}_{i j k l}(\omega) + ( \omega^2 + \Delta_{ik} \Delta_{jl} ) \tilde{I}^{\prime}_{i j k l}(\omega)\,.
\label{prime difference}
\end{equation}
In the case of two propagators, the following formula is useful to calculate the difference between the gauges,
%
%\begin{widetext}
\begin{align}
  I_{i_1 j_1 k_1 l_1}^{F}&(\omega_1) I_{i_2 j_2 k_2 l_2}^{F}(\omega_2) - I_{i_1 j_1 k_1 l_1}^{C}(\omega_1) I_{i_2 j_2 k_2 l_2}^{C}(\omega_2)  \nonumber\\ 
  = \frac{1}{2}&\left[ \Delta I_{i_1 j_1 k_1 l_1}(\omega_1) I_{i_2 j_2 k_2 l_2}^{F}(\omega_2)\right.\nonumber\\
  +& \left.I_{i_1 j_1 k_1 l_1}^{C}(\omega_1) \Delta I_{i_2 j_2 k_2 l_2}(\omega_2) + F \leftrightarrow C  \right]\,. 
\label{2propagator_difference}
\end{align}
%\end{widetext}
%
We note that it treats symmetrically both $I$-operators with respect to the gauge variation. It applies as well in the case of derivative terms, replacing $I_{i j k l}$ with $I^{\prime}_{i j k l}$. The following rearrangement formula will be handy to demonstrate the cancellation among terms of the three-electron part,
% . Assume $I(\boldsymbol{x} -\boldsymbol{y})$ and $I(\boldsymbol{z} -\boldsymbol{u})$ are hermitian operators and $[I(\boldsymbol{x} -\boldsymbol{y}), I(\boldsymbol{z} -\boldsymbol{u})]=0$, then based on the completeness relation one has
% %
% \begin{eqnarray}
% &&\sum_i  I_{j_1 l_1 i k_1} (\Delta_{k_1 l_1})   I_{j_2 i l_2 k_2} (\Delta_{j_2 l_2})  \nonumber \\ && =\sum_i  I_{k_2 k_1 i l_1}(\Delta_{l_1 k_1})  I_{l_2 i j_2 j_1}(\Delta_{l_2 j_2})\,.
% \label{re-arrangement}
% \end{eqnarray}
% %
%
\begin{align}
  \sum_m  I_{i_1 j_1 k_1 m} (\Delta_{k_1 i_1})   I_{m j_2 k_2 l_2} (\Delta_{j_2 l_2})  \nonumber \\
  = \sum_m  I_{j_2 j_1 l_2 m}(\Delta_{j_2 l_2})  I_{m i_1 k_2 k_1}(\Delta_{k_1 i_1})\,.
\label{re-arrangement}
\end{align}
It is based on the completeness relation and the commutation between the $I$-operators.
\subsection{Perturbation theory}
The interaction with the quantized electromagnetic field $A_{\mu}$ together with a {\em counterpotential} term are encapsulated in the interaction term
\begin{equation}
h_{\text{int}} = e \alpha^\mu A_\mu(x) - U(\boldsymbol{x})\,,
\end{equation}
with the corresponding normal-ordered interaction Hamiltonian
\begin{equation}
H_{\text{int}} = \int d^{3}x :\psi^{(0)\dagger}_\alpha(x) h_{\text{int}} \psi^{(0)}_\alpha(x):\,.   
\end{equation}
In this context the perturbation theory is based on the fine structure constant $\alpha$ as an expansion parameter, accounting for order by order corrections, assuming the $U(\boldsymbol{x})$ expansion converges reasonably fast,   
\begin{equation}
\Delta E_v = E_v -  E^{(0)}_v = \Delta E^{(1)}_v + \Delta E^{(2)}_v + ... 
\label{E pert exp}
\end{equation}
with the zero-order term, Eq.~(\ref{ezero}), being the energy obtained by solving the Dirac equation. In the original Furry picture, the $\alpha$ perturbative expansion can be enhanced with a further $1/Z$ expansion at each order. For an overall correction of order $n$, contributions range from $n{\text{th}}$ order in $\alpha$ up to $n{\text{th}}$ order in $1/Z$, with all possible intermediates fulfilling the order in $\alpha$ and in $1/Z$ corrections sum to $n$. The order in $1/Z$ indicates the number of exchanged photons, which mediates the interelectronic interaction between different electrons. In order to find the expressions for the energy shift (\ref{E pert exp}) the two-time Green's function approach \cite{shabaev:1990:43, shabaev:2002:119} is employed, which we formulate in the second quantization notations introduced above. Thus, the energy shift is defined as 
\begin{equation}
\Delta E_v = \frac{\displaystyle{\frac{1}{2\pi i}} \oint_{\Gamma_v} dE (E-\bra{v} H_0 \ket{v}) \bra{v}\Delta g_{\alpha}(E)\ket{v}} {1 + \displaystyle{\frac{1}{2\pi i}} \oint_{\Gamma_v} dE \bra{v}\Delta g_{\alpha}(E)\ket{v}}\,,
\label{energy correction}
\end{equation}
with $\Delta g_{\alpha}(E)= g_{\alpha}(E)- g_{\alpha}^{(0)}(E)$ and $\Gamma_v$ is the contour surrounding only the pole $E = \bra{v} H_0 \ket{v}$. The Fourier transformed two-time Green's function reads
\begin{eqnarray}
&&g_{\alpha}(E) \delta(E-E^{\prime}) = \frac{1}{2\pi i} \sum_{i,j}\int d^{3}x dx^0 dy^{0} e^{i (E x^{0} -E^{\prime}y^{ 0})} \nonumber\\
&&\times \phi^{\dagger}_i(\boldsymbol{x})\bra{\alpha} T(\psi_{\alpha} (x^{ 0}, \boldsymbol{x}) \psi_{\alpha}^{\dagger} (y^{0}, \boldsymbol{x}) ) \ket{\alpha}\phi_j(\boldsymbol{x})a_i^{\dagger}a_j\,.
\end{eqnarray}
The diagram technique stays the same as for the normal vacuum; the Feynman rules and detailed formulation of the two-time Green's function method can be found in Ref. \cite{shabaev:2002:119}. The perturbation theory is then formulated for the Green's function $g_{\alpha}(E)$:
\begin{equation}
\Delta g_{\alpha}(E) = \Delta g_{\alpha}^{(1)}(E) + \Delta g_{\alpha}^{(2)}(E) + ... \,.
\end{equation}
Separating out the corresponding order from Eq. (\ref{energy correction}), one gets the first- and second-order corrections to the energy shift
\begin{equation}
\Delta E^{(1)}_v = \frac{1}{2\pi i}\oint_{\Gamma_v} dE (E-\epsilon_v)\Delta g_{\alpha,vv}^{(1)}(E)\,, 
\label{1st O corr}
\end{equation}
and
\begin{eqnarray}
\Delta E^{(2)}_v &=& \frac{1}{2\pi i}\oint_{\Gamma_v} dE (E-\epsilon_v) \Delta g_{\alpha,vv}^{(2)}(E)\nonumber\\
&-&\frac{1}{2\pi i}\oint_{\Gamma_v} dE (E-\epsilon_v) \Delta g_{\alpha,vv}^{(1)}(E)\nonumber\\
&\times&\frac{1}{2\pi i} \oint_{\Gamma_v} dE^{\prime} \Delta g_{\alpha,vv}^{(1)}(E^{\prime})\,,
\label{2nd O cor}
\end{eqnarray}
where $\Delta g_{\alpha,vv}(E) = \bra{v}\Delta g_{\alpha}(E)\ket{v}$. The vacuum redefinition allows us to consider only one-electron Feynman diagrams, which, however, contain also the many-electron contributions. A similar method was previously employed in  the evaluation of the screened radiative and two-photon exchange corrections to the $g$ factor and hyperfine splitting \cite{volotka:2009:033005, glazov:2010:062112, volotka:2012:073001, volotka:2014:253004, glazov:2019:173001, kosheleva:2020:013364}, however, it was not described in detail. Here, we investigate how the interelectronic-interaction corrections, namely, one- and two-photon exchange, to binding energies can be extracted from the one-electron diagrams. To make it clear, starting from the first-order correction in $\alpha$, a subdivision among different contributions of the same order is made. The one-electron loop diagrams are denoted by $(\text{L})$, the interelectronic-interaction diagrams by $(\text{I})$ and the screened loop diagrams by $(\text{S})$. The latter will only start to occur to the second order. The first-order correction,
\begin{equation}
\Delta E^{(1)}_v = \Delta E^{(1\text{L})}_v + \Delta E^{(1\text{I})}_v\,, 
\label{1stcorrection}
\end{equation}
contains the self-energy (SE) and vacuum-polarization (VP) contributions in the first term and the counterpotential (CP) as well as the one-photon exchange in the second one. The second-order correction,
\begin{equation}
\Delta E^{(2)}_v = \Delta E^{(2\text{L})}_v + \Delta E^{(2\text{I})}_v + \Delta E^{(2\text{S})}_v\,,
\end{equation}
provides the one-electron two-loop contributions, the two-photon exchange and the screened one-loop corrections in the respective order. The focus of the current paper is the two-photon exchange term $\Delta E^{(2\text{I})}_v$. In the next section, we demonstrate how to extract the interelectronic-interaction correction on an example of the one-photon exchange.

%=================================================================
%
\section{One-photon exchange}
\begin{figure}
\centering
\includegraphics[scale=0.5]{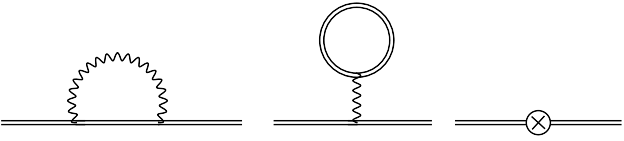}
\caption{The first-order one-electron Feynman diagrams corresponding (from left to right) to SE, VP, and CP corrections. Double lines indicate electron propagators in the external potential $V$. Wavy lines correspond to the photon propagator. The cross inside a circle represents a counterpotential term, $-U$.}
\label{fig:1}
\end{figure} 
In this section, the method and techniques presented above will be illustrated by an example of first-order diagrams, which are depicted in Fig. \ref{fig:1}. The matrix elements of the Green's function for SE and VP are, using the Feynman rules provided in \cite{shabaev:2002:119}, respectively, given by
\begin{align}
\Delta g^{(1)\text{SE}}_{\alpha,vv}(E) =& \frac{1}{(E - \epsilon_v)^2} \frac{i}{2\pi} \int d\omega 
\nonumber\\
&\sum_{j} \frac{I_{v j j v}(\omega)  }{E -\omega -\epsilon_{j} + i\varepsilon (\epsilon_j-E_{\alpha}^{F})}\,,
\\
\Delta g^{(1)\text{VP}}_{\alpha,vv}(E) =& \frac{-1}{(E - \epsilon_v)^2} \frac{i}{2\pi} \int d\omega 
\nonumber\\
&\sum_{j} \frac{I_{v j v j}(0)}{\omega -\epsilon_{j} + i\varepsilon (\epsilon_j-E_{\alpha}^{F})}\,.
\end{align} 
They are used as inputs to calculate the correction to the energy according to Eq.~(\ref{1st O corr}). The contour integral contains a first order pole in $E = \epsilon_v$, thus the Green's functions are evaluated at $E= \epsilon_v$ leading to 
\begin{align}
\Delta E^{(1)\text{SE}}_v =& \frac{i}{2\pi} \int d\omega \sum_{j} \frac{I_{v j j v}(\omega)  }{\epsilon_v -\omega -\epsilon_{j} + i\varepsilon (\epsilon_j-E_{\alpha}^{F})}\,,
\\
\Delta E^{(1)\text{VP}}_v =&  - \frac{i}{2\pi} \int d\omega \sum_{j} \frac{I_{v j v j}(0)}{\omega -\epsilon_{j} + i\varepsilon (\epsilon_j-E_{\alpha}^{F})}\,.
\end{align} 
Hence, one got the first-order SE and VP energy corrections in the redefined vacuum framework. We notice that these formulas encapsulate not only the one-electron one-loop corrections, but also the one-photon exchange. According to Eq. (\ref{1stcorrection}) , in order to obtain the one-photon-exchange contribution, one should subtract SE and VP corrections calculated in the framework of the usual vacuum. With the help of Eq.~(\ref{Sokhotski}) we obtain for the SE part,
\begin{align}
  \Delta E^{(1\text{I})\text{SE}}_v &= \Delta E^{(1)\text{SE}}_v - \Delta E^{(1\text{L})\text{SE}}_v  
\nonumber\\
  &= \frac{i}{2\pi} \int d\omega \sum_{j} \left[ \frac{I_{v j j v}(\omega)}{\epsilon_v -\omega -\epsilon_{j} + i\varepsilon (\epsilon_j-E_{\alpha}^{F})} 
  \right.
\nonumber\\
  &\left.
  - \frac{I_{v j j v}(\omega)}{\epsilon_v -\omega -\epsilon_{j}u} \right] 
  = - \sum_{a} I_{v a a v}(\Delta_{va})
\,,
\end{align}
and for the VP part,
\begin{align}
  \Delta E^{(1\text{I})\text{VP}}_v &=  \Delta E^{(1)\text{VP}}_v -  \Delta E^{(1\text{L})\text{VP}}_v
\nonumber\\ 
  &= -\frac{i}{2\pi} \int d\omega \sum_{j} \left[ \frac{I_{v j v j}(0)}{\omega -\epsilon_{j} + i\varepsilon (\epsilon_j-E_{\alpha}^{F})} 
  \right.
\nonumber\\
  &\left.
  - \frac{I_{v j v j}(0)}{\omega -\epsilon_{j} u} \right]
  = \sum_{a} I_{v a v a}(0)
\,.
\end{align}
It is left to evaluate the counterpotential graph. Let us define $\int d^{3}x    \phi_i^{\dagger}(\boldsymbol{x}) U(\boldsymbol{x})\phi_j(\boldsymbol{x}) \equiv U_{ij}$, then the corresponding Green's function takes the form 
\begin{equation}
\Delta g^{(1)\text{CP}}_{\alpha,vv}(E) = \frac{-U_{v  v}}{(E - \epsilon_v)^2}\,. 
\end{equation}
Following the same line as before, the evaluation of the contour integral in Eq. (\ref{1st O corr}) provides
\begin{equation}
\Delta E^{(1\text{I})\text{CP}}_v =  \Delta E^{(1)\text{CP}}_v = -U_{vv}\,, 
\end{equation}
since CP does not participate in radiative corrections belonging to $\Delta E^{(1\text{L})}_v$. The total first order correction $\Delta E^{(1\text{I})}_v $ is found to be
\begin{equation}
\Delta E^{(1\text{I})}_v = \sum_{a} \left[I_{vava}(0) - I_{vaav}(\Delta_{va})\right] - U_{vv}\,.
\end{equation}
Let proceed to demonstrate the gauge invariance of the first order correction $\Delta E^{(1\text{I})}_v$. The counterpotential term is obviously  gauge invariant. The starting point for the two remaining terms is Eq. (\ref{propagator_difference}). Straightforward application of this formula gives
\begin{eqnarray}
\delta \Delta E^{(1\text{I})}_v &=& \sum_a \left[(0^2 + \Delta_{vv}\Delta_{cc}) \tilde{I}_{vava}(0)\right.\nonumber\\ &-& \left.(\Delta_{va}^2 + \Delta_{va} \Delta_{av}) \tilde{I}_{v a a v}(\Delta_{va})\right] = 0\,,
\end{eqnarray}
thus, showing its gauge invariance. 

%=================================================================
%
\section{Two-photon exchange diagrams and gauge invariant subsets}
\begin{figure}
\centering
\includegraphics[width=0.47\textwidth]{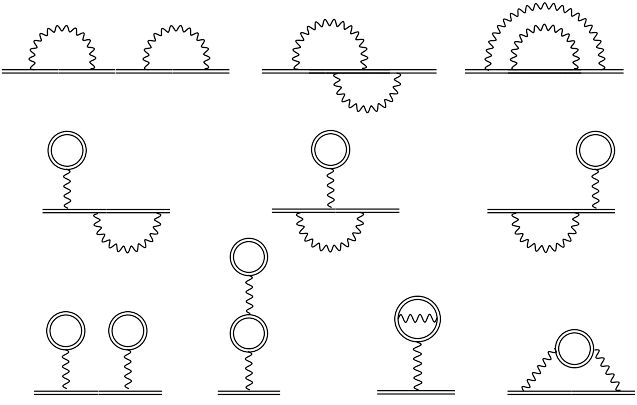}
\caption{The second-order one-electron Feynman diagrams labeled as follows: SESE (first row); SEVP (second row); VPVP, V(VP)P, V(SE)P, and S(VP)E from left to right in the last row. Other notations are the same as in Fig. \ref{fig:1}.}
\label{fig:2}
\end{figure} 
\begin{figure}
\centering
\includegraphics[width=0.47\textwidth]{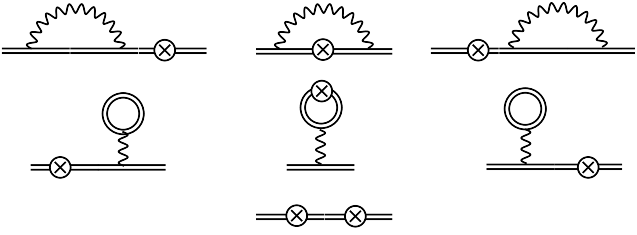}
\caption{The second-order one-electron counterpotential Feynman diagrams labeled as follows: SECP (first row); VPCP (second row); CPCP (third row). Other notations are the same as in Fig. \ref{fig:1}.}
\label{fig:3}
\end{figure}
In order to derive the expressions of the matrix elements for the two-photon exchange corrections in the redefined vacuum framework, one has to start with the complete set of second-order one-electron diagrams. There are ten two-loop diagrams, which are presented in Fig. \ref{fig:2}. In the extended Furry picture, one has to consider also seven counterpotential diagrams depicted in Fig. \ref{fig:3}. Already such a decomposition allows us to identify nine gauge invariant subsets based on the gauge invariance of the one-electron two-loop diagrams \cite{yerokhin:2006:253004}. Here are the subsets with labeling presented in Figs. \ref{fig:2} and \ref{fig:3}: SESE, SEVP, VPVP, V(VP)P, V(SE)P, S(VP)E, SECP, VPCP, and CPCP. The identified subsets should be gauge invariant in both redefined and usual vacuum frameworks. This means that the many-electron diagrams obtained as a difference between redefined and usual vacuum diagrams can be also divided according to these subsets. In the following sections this statement will be illustrated by the rigorous derivation and direct proof of the gauge invariance.

Later, a distinction has to be made between \emph{reducible} and \emph{irreducible} contributions. The first corresponds to the intermediate energy of the system being equal to its initial energy, and the latter is the remaining part. To keep track of the source of generated reducible contributions, a subscript is used with previous notation; for example $v_1, a_1, b_1$, where $\epsilon_{i_1}= \epsilon_i$.
\begin{widetext}
%
%---------------------------------
%
\subsection{SESE subset}
The Green's function for the SESE subset can be cast in the form
\begin{align}
\Delta g^{(2)\text{SESE}}_{\alpha,vv}(E) &= \frac{1}{(E - \epsilon_v)^2} \left(\frac{i}{2\pi}\right)^2 \sum_{i, j, k} \int d\omega_1 d\omega_2
\nonumber\\
&\times\left\{
\frac{I_{v i i j}(\omega_1) I_{j k k v}(\omega_2)}
{[ E - \omega_1 - \epsilon_i + i\varepsilon (\epsilon_i-E_\alpha^F) ]
 [ E - \epsilon_j + i\varepsilon (\epsilon_j-E_\alpha^F) ]
 [ E - \omega_2 - \epsilon_k + i\varepsilon (\epsilon_k-E_\alpha^F) ]}
\right.
\nonumber\\
&+
\frac{I_{v j i k}(\omega_1) I_{i k j v}(\omega_2)}
{[ E - \omega_1 - \epsilon_i + i\varepsilon (\epsilon_i-E_\alpha^F) ]
 [ E - \omega_1 - \omega_2 - \epsilon_j + i\varepsilon (\epsilon_j-E_\alpha^F) ]
 [ E - \omega_2 - \epsilon_k + i\varepsilon (\epsilon_k-E_\alpha^F) ]}
\nonumber\\
&+\left.
\frac{I_{v k i v}(\omega_1) I_{i j j k}(\omega_2)}
{[ E - \omega_1 - \epsilon_i + i\varepsilon (\epsilon_i-E_\alpha^F) ]
 [ E - \omega_1 - \omega_2 - \epsilon_j + i\varepsilon (\epsilon_j-E_\alpha^F) ]
 [ E - \omega_1 - \epsilon_k + i\varepsilon (\epsilon_k-E_\alpha^F) ]}\right\}\,,
\label{eq:gSESE}
\end{align}
with the corresponding energy correction, see Eq.~(\ref{2nd O cor}),
\begin{equation}
\label{2ndoSESE}
\Delta E^{(2)\text{SESE}}_v = \frac{1}{2\pi i} \oint_{\Gamma_v} dE (E - \epsilon_v) \Delta g_{\alpha,vv}^{(2)\text{SESE}}(E) - \frac{1}{2\pi i}\oint_{\Gamma_v} dE (E - \epsilon_v) \Delta g_{\alpha,vv}^{(1)\text{SE}}(E) \frac{1}{2\pi i}\oint_{\Gamma_v} dE^{\prime}\Delta g_{\alpha,vv}^{(1)\text{SE}}(E^{\prime})\,.
\end{equation}
Let us now work through the two-photon exchange extraction procedure,
\begin{equation}
\Delta E^{(2)\text{SESE}}_v - \Delta E^{(2\text{L})\text{SESE}}_v = \Delta E^{(2\text{I})\text{SESE}}_v + \Delta E^{(2\text{S})\text{SESE}}_v\,,
\end{equation}
for the last term of $\Delta g^{(2)\text{SESE}}_{\alpha,vv}(E)$ in Eq.~(\ref{eq:gSESE}) as an example. In order to not overload the following steps, the integrals and associated prefactors are ignored for the time being, however, signs are taken into account accordingly:
\begin{align}
\label{eq:extraction}
&\sum_{i, j, k}\left\{\frac{I_{v k i v}(\omega_1)  I_{i j j k}(\omega_2)}
{[ E - \omega_1 - \epsilon_i + i\varepsilon (\epsilon_i-E_\alpha^F) ]
 [ E - \omega_1 - \omega_2 - \epsilon_j + i\varepsilon (\epsilon_j-E_\alpha^F) ]
 [ E - \omega_1 - \epsilon_k + i\varepsilon (\epsilon_k-E_\alpha^F) ]}\right.
\nonumber\\
&\hspace{9cm}-\left.\frac{I_{v k i v}(\omega_1) I_{i j j k}(\omega_2)}
{[ E - \omega_1 - \epsilon_{i} u]
 [ E - \omega_1 - \omega_2- \epsilon_{j} u ]
 [ E - \omega_1 - \epsilon_{k} u ]}\right\}
\nonumber\\
&= 2\pi i \sum_{a,j,k}^{k \neq a} \frac{I_{v k a v}(\omega_1)  I_{a j j k}(\omega_2)\delta(E - \omega_1 - \epsilon_{a})}{ (E - \omega_1 - \omega_2- \epsilon_{j} u) (E - \omega_1 - \epsilon_{k } u)}
+ 2\pi i \sum_{i,a,k} \frac{ I_{v k i v}(\omega_1)  I_{i a a k}(\omega_2) \delta (E - \omega_1 - \omega_2- \epsilon_{a})}{(E - \omega_1 - \epsilon_{i} u) (E - \omega_1 - \epsilon_{k} u)}\nonumber\\
&+ 2\pi i \sum_{i, j, a}^{i \neq a}  \frac{ I_{v a i v}(\omega_1)  I_{i j j a}(\omega_2) \delta (E - \omega_1 - \epsilon_{a} )}{(E - \omega_1 - \epsilon_{i} u) (E - \omega_1 - \omega_2- \epsilon_{j} u)}
+ (2\pi i)^2 \sum_{a,b,k}^{k \neq a} \frac{ I_{v k a v}(\omega_1) I_{a b b k}(\omega_2) \delta(E - \omega_1 - \epsilon_{a}) \delta(E - \omega_1 - \omega_2- \epsilon_{b} )}{  (E - \omega_1 - \epsilon_{k } u)} \nonumber\\
&+ (2\pi i)^2 \sum_{a,j,a_1} \frac{ I_{v a_1 a v}(\omega_1)  I_{a j j a_1}(\omega_2)    }{ (E - \omega_1 - \omega_2- \epsilon_{j} u) } \frac{d}{d\omega_1}\delta(E - \omega_1 - \epsilon_a)\nonumber\\
&+ (2\pi i)^2 \sum_{i,a,b}^{i \neq b} \frac{ I_{v b i v}(\omega_1)  I_{i a a b}(\omega_2) \delta (E - \omega_1 - \omega_2- \epsilon_{a}) \delta(E - \omega_1 - \epsilon_{b} )}{(E - \omega_1 - \epsilon_{i} u) }\nonumber\\
&+ (2\pi i)^2 \sum_{a,b,a_1}  I_{v a_1 a v}(\omega_1)  I_{a b b a_1}(\omega_2)  \delta(E - \omega_1 - \omega_2- \epsilon_{b} )    \frac{d}{d\omega_1}\delta(E - \omega_1 - \epsilon_a)\,.
\end{align}
After one cut, one gets the first three terms with only one delta function. The first and third of them correspond to the screened radiative diagrams, while the second one gives the crossed diagram of the two-photon exchange. At the two-cut level, two three-electron contributions are found, the fourth and sixth terms with two delta functions. The term which has a derivative acting on the delta function gives rise to the reducible contribution of the screened radiative diagram. The reducible three-electron part is provided by the three-cut term displayed on the last line of Eq. (\ref{eq:extraction}). Thus, separating only the interelectronic-interaction terms, we get for the last term of Eq.~(\ref{eq:gSESE}):
\begin{eqnarray}
\Delta E^{(2\text{I})\text{SESE}}_v &\sim& 2\pi i \sum_{i,a,k} \frac{ I_{v k i v}(\omega_1)  I_{i a a k}(\omega_2) \delta (E - \omega_1 - \omega_2- \epsilon_{a})}{(E - \omega_1 - \epsilon_{i} u) (E - \omega_1 - \epsilon_{k} u)}\nonumber\\
&+& (2\pi i)^2 \sum_{a,b,k}^{k \neq a} \frac{ I_{v k a v}(\omega_1) I_{a b b k}(\omega_2) \delta(E - \omega_1 - \epsilon_{a}) \delta(E - \omega_1 - \omega_2- \epsilon_{b} )}{  (E - \omega_1 - \epsilon_{k } u)}  \nonumber\\
&+& (2\pi i)^2 \sum_{i,a,b}^{i \neq b}  \frac{ I_{v b i v}(\omega_1)  I_{i a a b}(\omega_2) \delta (E - \omega_1 - \omega_2- \epsilon_{a}) \delta(E - \omega_1 - \epsilon_{b} )}{(E - \omega_1 - \epsilon_{i} u) } \nonumber\\
&+& (2\pi i)^2\sum_{a,b,a_1}  I_{v a_1 a v}(\omega_1)  I_{a b b a_1}(\omega_2)  \delta(E - \omega_1 - \omega_2- \epsilon_{b} ) \frac{d}{d\omega_1}\delta(E - \omega_1 - \epsilon_{a_1}) + ...\,.
\end{eqnarray}
Notice that one obtains different types of contribution arising when the same diagram in the usual vacuum is subtracted: two-electron and three-electron, which are additionally divided into irreducible and reducible parts. The disconnected SESE contribution, the second term of Eq. (\ref{2ndoSESE}), is incorporated into the three-electron reducible part. 

In the rest of the section we summarize the results for the SESE contribution.
%
%\subsubsection{Two-electron SESE terms}
%
The two-electron irreducible part $\Delta E^{(2\text{I})\text{SESE,2e,irr}}_{v}$ is given by
\begin{eqnarray}
\Delta E^{(2\text{\text{I}})\text{SESE,2e,irr}}_{v} =-\frac{i}{2\pi} \int d\omega \left[ \sum_{a, i, j}^\prime \frac{I_{v j i v}(\omega) I_{i a a j}(\Delta_{va}-\omega)}{(\epsilon_v - \omega - \epsilon_{i} u)  (\epsilon_v - \omega - \epsilon_{j} u)} + \sum_{a,i,j}^{(i,j) \neq (a,v)} \frac{ I_{v a i j}(\omega)  I_{i j a v}(\Delta_{va}-\omega)    }{(\epsilon_v - \omega - \epsilon_{i} u)  (\epsilon_a + \omega - \epsilon_{j} u)} \right]\,,
\label{eq:SESE,2e,irr}
\end{eqnarray}
where the first term resembles the crossed exchange graph considered above in details, while the second one gives the ladder exchange diagram. Following Ref. \cite{yerokhin:2001:032109}, we exclude (prime on the sum) from the crossed irreducible term the contributions with $\epsilon_{i} + \epsilon_{j} = 2\epsilon_v$ and $\epsilon_{i} + \epsilon_{j} = 2\epsilon_a$ and attribute them to the reducible term. The two-electron reducible part $\Delta E^{(2\text{I})\text{SESE,2e,red}}_{v}$, which corresponds to $\epsilon_{i} + \epsilon_{j} = \epsilon_a + \epsilon_v$ restriction in the ladder term and the contributions excluded in the irreducible crossed term, reads
\begin{eqnarray}
\Delta E^{(2\text{I})\text{SESE,2e,red}}_{v} &=& \frac{i}{2\pi}  \int \frac{d\omega}{(\omega + i\varepsilon)^2}\left\{ \sum_{a, a_1, v_1}  \left[ I_{v a v_1 a_1}(\omega)  I_{v_1 a_1 a v}(\Delta_{va}+\omega)  +  I_{v a a_1 v_1}(\Delta_{va}-\omega) I_{a_1 v_1 a v}(\omega) \right]\right.\nonumber\\
&-&\left.
   \sum_{a, v_1, v_2} I_{v v_2 v_1 v}(\omega) I_{v_1 a a v_2}(\Delta_{va}+\omega)
 - \sum_{a, a_1, a_2} I_{v a_2 a_1 v}(\Delta_{va}-\omega) I_{a_1 a a a_2}(\omega) \right\}\,.
\label{eq:SESE,2e,red}
\end{eqnarray} 
%
%\subsubsection{Three-electron SESE terms}
%
The three-electron irreducible part $\Delta E^{(2\text{I}) \text{SESE,3e,irr}}_{v}$ is given by
\begin{eqnarray}
\Delta E^{(2\text{I})\text{SESE,3e,irr}}_{v} &=& \sum_{a,b,i}^{i \neq b}  \frac{ I_{v i b v}(\Delta_{vb})  I_{b  a a  i}(\Delta_{ba})  + I_{v b i v}(\Delta_{vb})  I_{i  a a  b}(\Delta_{ab})    }{(\epsilon_b -\epsilon_{i} ) } + \sum_{a,b,i}^{i \neq v}  \frac{ I_{v a a i}(\Delta_{va})  I_{i b b   v}(\Delta_{vb})    }{(\epsilon_v - \epsilon_{i} ) } \nonumber \\ &+&\sum_{a,b,i}^{(i, b) \neq (v,a)}  \frac{ I_{v a b i}(\Delta_{vb})  I_{b i a v }(\Delta_{ba})  +  I_{v a i b }(\Delta_{ba})  I_{i b a v }(\Delta_{vb})    }{(\epsilon_v+ \epsilon_a-\epsilon_b-\epsilon_{i} ) } + \sum_{a,b,i}\frac{ I_{v i a b }(\Delta_{va})  I_{a b i v }(\Delta_{vb})    }{(\epsilon_a + \epsilon_b -\epsilon_v - \epsilon_{i} ) }\,.
\label{eq:SESE,3e,irr}
\end{eqnarray}
The derivative term arising from the ladder exchange graph as well as the disconnected SESE contribution from the second term in Eq. (\ref{2nd O cor}) are merged into the reducible part, leading to
\begin{eqnarray}
\Delta E^{(2\text{I})\text{SESE,3e,red}}_{v} &=&  \sum_{a, b, a_1}
\left[ I_{v a_1 a v}(\Delta_{va}) I^{\prime}_{a b b a_1}(\Delta_{ab})
     - I^{\prime}_{v a_1 a v}(\Delta_{va}) I_{a b b a_1}(\Delta_{ab})\right]
\nonumber\\
&+& \sum_{a, b, v_1} I^{\prime}_{v a a v_1}(\Delta_{va}) I_{v_1 b b v}(\Delta_{vb})
 +  \sum_{a, a_1, v_1} I^{\prime}_{v a a_1 v_1 }(\Delta_{va}) I_{a_1 v_1 a v}(0)\,.
\label{eq:SESE,3e,red}
\end{eqnarray}
%
%\subsubsection{Gauge invariance of the three-electron SESE subset}
%
Since the contributions summarized above are derived from the gauge invariant SESE subset, we assume that they form the gauge invariant subset as well. Moreover, the gauge invariance also persists for two- and three-electron parts independently. The gauge invariance of the three-electron SESE subset can be demonstrated analytically. With the help of Eq. (\ref{2propagator_difference}), once the denominators have been canceled and the sum completed with the appropriated missing terms, the gauge difference $\delta \Delta E^{(2),\text{SESE,3e,irr}}_{v}$ is turned into
\begin{eqnarray}
&&\delta\Delta E^{(2\text{I})\text{SESE,3e,irr}}_{v} = \frac{1}{2} \sum_{a,b,i}
\left[
    \Delta_{bv} \tilde{I}_{v i b v}(\Delta_{vb}) I^{F}_{b a a i}(\Delta_{ba})
  + \Delta_{ba} I^{C}_{v i b v}(\Delta_{vb}) \tilde{I}_{b a a i}(\Delta_{ba})
  + \Delta_{vb} \tilde{I}_{i b a v}(\Delta_{vb}) I^{F}_{v a  i b}(\Delta_{ab})
\right.\nonumber\\
&&+ \Delta_{ab} I^{C}_{i b a v}(\Delta_{vb}) \tilde{I}_{v a i b}(\Delta_{ab}) 
  + \Delta_{bv} \tilde{I}_{v b i v}(\Delta_{vb}) I^{F}_{i a a b}(\Delta_{ab})
  + \Delta_{ba} I^{C}_{v b i v}(\Delta_{vb}) \tilde{I}_{i a a b}(\Delta_{ab})
  + \Delta_{vb} \tilde{I}_{v a b i}(\Delta_{vb}) I^{F}_{b i a v}(\Delta_{ba})
\nonumber\\  
&&+ \Delta_{ab} I^{C}_{v a b i}(\Delta_{vb}) \tilde{I}_{b i a v}(\Delta_{ba}) 
  + \Delta_{va} \tilde{I}_{v a a i}(\Delta_{va}) I^{F}_{i b b v}(\Delta_{vb})
  + \Delta_{vb} I^{C}_{v a a i}(\Delta_{va}) \tilde{I}_{i b b v}(\Delta_{vb})
  + \Delta_{av} \tilde{I}_{v i a b}(\Delta_{va}) I^{F}_{a b i v}(\Delta_{vb})
\nonumber\\
&&\left.
  + \Delta_{bv} I^{C}_{v i a b}(\Delta_{va}) \tilde{I}_{a b i v}(\Delta_{vb})
  \right]
+ F \leftrightarrow C - \delta^{\text{SESE}}_{\text{completion}}\,.
\end{eqnarray}
Applying then Eq. (\ref{re-arrangement}) on the 1st and 2nd terms with appropriate use of symmetry property (\ref{eq:symmetry}) allows us to rewrite them exactly as the 3rd and 4th terms but with an opposite sign, therefore, canceling them. And doing similar for other terms shows the absence of contribution coming from the complete sum. The remaining terms arising from the completion of the sum are given by the expression
\begin{eqnarray}
\delta^{\text{SESE}}_{\text{completion}} &=& \sum_{a, b, a_1} \left[
 \Delta_{av} \tilde{I}_{v a_1 a v}(\Delta_{va}) I^{F}_{a b b a_1}(\Delta_{ab})
+\Delta_{ab} I^{C}_{v a_1 a v}(\Delta_{va}) \tilde{I}_{a b b a_1}(\Delta_{ab})\right]
+\sum_{a, b, v_1} \Delta_{vb} \tilde{I}_{v a b v_1}(\Delta_{vb}) I^{F}_{b v_1 a v}(0)
\nonumber\\
&+&\frac12\sum_{a, b, v_1}\left[
  \Delta_{va} \tilde{I}_{v a a v_1}(\Delta_{va}) I^{F}_{v_1 b b v}(\Delta_{vb})
+ \Delta_{vb} I^{C}_{v a a v_1}(\Delta_{va}) \tilde{I}_{v_1 b b v}(\Delta_{vb})\right]
+ F \leftrightarrow C\,.
\end{eqnarray}
The contribution from the reducible part reads
\begin{eqnarray}
\delta\Delta E^{(2\text{I})\text{SESE,3e,red}}_{v} &=& \sum_{a, b, a_1} \left[
 \Delta_{av} \tilde{I}_{v a_1 a v}(\Delta_{va}) I^{F}_{a b b a_1}(\Delta_{ab})
+\Delta_{ab} I^{C}_{v a_1 a v}(\Delta_{va}) \tilde{I}_{a b b a_1}(\Delta_{ab})\right] 
\nonumber\\
&+&\sum_{a, b, v_1}\left[
\Delta_{vb} \tilde{I}_{v a b v_1}(\Delta_{vb}) I^{F}_{b v_1 a v}(0) + \Delta_{va} \tilde{I}_{v a a v_1}(\Delta_{va}) I^{F}_{v_1 b b v}(\Delta_{vb})\right] + F \leftrightarrow C\,.
\end{eqnarray}
One can see that the first three terms of the completion term cancel out the first three terms of the reducible part. The remaining ones vanish when their counterparts in $F \leftrightarrow C$ are taken into account, which leads to the desired result.
%
%---------------------------------
%
\subsection{SEVP subset}
The SEVP subset contains only three-electron contributions, which give the following irreducible part $\Delta E^{(2\text{I})\text{SEVP,3e,irr}}_{v}$,
\begin{equation}
\Delta E^{(2\text{I})\text{SEVP,3e,irr}}_{v} =   -\sum_{a,b,i}^{i \neq v}  \frac{ I_{v a a i}(\Delta_{va})  I_{i b v b}(0) +  I_{i b b v}(\Delta_{vb}) I_{v a i a }(0)  }{(\epsilon_v -\epsilon_{i} ) }  -\sum_{a,b,i}^{i \neq a}  \frac{ I_{v i a v}(\Delta_{va})  I_{a b i b}(0) +I_{v a i v}(\Delta_{va})  I_{i b a b}(0)   }{(\epsilon_a -\epsilon_{i} ) }\,,
\label{eq:SEVP,3e,irr}
\end{equation}
and the reducible part being merged with the disconnected SEVP contribution yields
\begin{equation}
\Delta E^{(2\text{I}) \text{SEVP,3e,red}}_{v} =
- \sum_{a, b, v_1} I^{\prime}_{v a a v_1}(\Delta_{va})  I_{v_1 b v b}(0)
+ \sum_{a, b, a_1} I^{\prime} _{v a_1 a v}(\Delta_{va})  I_{a b a_1 b}(0)\,.  
\label{eq:SEVP,3e,red}
\end{equation}
In order to show its gauge invariance, the first thing one has to do after using Eq. (\ref{2propagator_difference}) and canceling the denominator is to complete the sum. These extra terms cancel the reducible part after simple $a \leftrightarrow b$ renaming where necessary. One is left with
\begin{eqnarray}
\delta\Delta E^{(2\text{I})\text{SEVP,3e,irr}}_{v} &=& -\frac{1}{2} \sum_{a,b,i} \left[ \Delta_{va} \tilde{I}_{v a a n}(\Delta_{va}) I^{F}_{n b v b}(0) + \Delta_{vb} \tilde{I}_{n b b v}(\Delta_{vb}) I ^{F}_{v a n a}(0) + \Delta_{av} \tilde{I}_{v n a v }(\Delta_{va}) I^{F}_{a b n b}(0)\right. \nonumber\\ &+& \left.\Delta_{av} \tilde{I}_{v a n v}(\Delta_{va}) I^{F}_{n b a b}(0)  + F \leftrightarrow C\right]\,.    
\end{eqnarray}
Applying Eq. (\ref{re-arrangement}) on the two first terms and after minor rewriting based on the symmetry properties, it gives exactly the same terms as the two last ones but with an opposite sign. Thus, the complete sum gives zero contribution and the subset is gauge invariant.
%
%---------------------------------
%
\subsection{S(VP)E subset}
Two-electron contributions are also found in the S(VP)E subset, in addition to the three-electron ones. Let us now consider them separately and start with the first ones.
%
%\subsubsection{Two-electron S(VP)E terms}
%
The two-electron irreducible part $\Delta E^{(2\text{I})\text{S(VP)E,2e,irr}}_{v}$ reads
\begin{equation}
\Delta E^{(2\text{I})\text{S(VP)E,2e,irr}}_{v} = \frac{i}{2\pi}\int d\omega \left[  \sum_{a, i, j}^\prime \frac{I_{v j i a}(\omega) I_{i a v j}( \omega)}{(\epsilon_v- \omega - \epsilon_{i} u) (\epsilon_a - \omega - \epsilon_{j} u)} + \sum_{a, i, j}^{(i,j) \neq (a,v)} \frac{ I_{v a i j }(\omega ) I_{i j v a}( \omega)}{(\epsilon_v- \omega - \epsilon_{i} u) (\epsilon_a + \omega - \epsilon_{j} u)} \right]\,,
\label{eq:S(VP)E,2e,irr}
\end{equation}
where we again exclude the contribution with $\epsilon_i = \epsilon_v$ and $\epsilon_j = \epsilon_a$ from the crossed direct term (first item) by hand and note this by the prime on the sum. The two-electron reducible part $\Delta E^{(2\text{I})\text{S(VP)E,2e,red}}_{v}$, coming from the ladder direct restriction $\epsilon_{i} + \epsilon_{j} = \epsilon_a + \epsilon_v$ together with the excluded contribution from the irreducible crossed direct term, yields
\begin{equation}
\Delta E^{(2\text{I})\text{S(VP)E,2e,red}}_{v} = -\frac{i}{4\pi} \int d\omega \sum_{a, a_1, v_1} I_{v a a_1 v_1 }(\Delta_{va}-\omega) I_{a_1 v_1 v a}(\Delta_{va}-\omega) \left[ \frac{1}{(\omega + i \varepsilon)^2}+ \frac{1}{(-\omega + i \varepsilon)^2} \right]\,.  
\label{eq:S(VP)E,2e,red}
\end{equation}
%
%\subsubsection{Three-electron S(VP)E terms}
%
The three-electron irreducible part $\Delta E^{(2\text{I})\text{S(VP)E,3e,irr}}_{v}$ yields
\begin{equation} 
\Delta E^{(2\text{I})\text{S(VP)E,3e,irr}}_{v} = -\sum_{a,b,i}^{(i, b) \neq (v, a)}  \frac{ I_{v a b i}(\Delta_{vb})  I_{b i v a}(\Delta_{vb})  + I_{v a i b}(\Delta_{ba})  I_{i b v a}(\Delta_{ba})  }{(\epsilon_v +\epsilon_a -\epsilon_b -\epsilon_{i} ) }  -\sum_{a,b,i}  \frac{ I_{v i b a}(\Delta_{vb})  I_{b a v i}(\Delta_{vb})    }{(\epsilon_a +\epsilon_b -\epsilon_v -\epsilon_{i} )}\,,
\label{eq:S(VP)E,3e,irr}
\end{equation}
together with the corresponding reducible part, 
\begin{equation}
\Delta E^{(2\text{I}) \text{S(VP)E,3e,red}}_{v} = -\sum_{a, a_1, v_1}
I_{v a a_1 v_1}(\Delta_{va}) I^{\prime}_{a_1 v_1 v a}(\Delta_{va})\,. 
\label{eq:S(VP)E,3e,red}
\end{equation}
%
%\subsubsection{Gauge invariance of the three-electron S(VP)E subset}
%
The recipe to show gauge invariance is always the same: apply Eq. (\ref{2propagator_difference}), cancel the denominators, and complete the sum. One obtains afterward for the three-electron part
\begin{eqnarray}
&&\delta \Delta E^{(2\text{I})\text{S(VP)E,3e,irr}}_{v} = -\frac{1}{2} \sum_{a,b,i} \left\{  \Delta_{bv} \left[ \tilde{I}_{v i b a}(\Delta_{vb}) I^{F}_{b a v i }(\Delta_{vb}) + I^{C}_{v i b a}(\Delta_{vb}) \tilde{I}_{b a v i} (\Delta_{vb}) \right] + \Delta_{vb}\left[ \tilde{I}_{v a b i}(\Delta_{vb}) I^{F}_{b i v a}(\Delta_{vb})\right.\right.\nonumber\\
&&+\left.\left. I^{C}_{v a b i}(\Delta_{vb}) \tilde{I}_{b i v a} (\Delta_{vb})\right]
+ \Delta_{ab}\left[ \tilde{I}_{v a i b }(\Delta_{ba}) I^{F}_{i b v a }(\Delta_{ba}) + I^{C}_{v a i b }(\Delta_{ba}) \tilde{I}_{i b v a} (\Delta_{ba}) \right]\right\} + F \leftrightarrow C -\delta^{\text{S(VP)E}}_{\text{completion}}\,.
\end{eqnarray}
The first two terms of the sum are rewritten with the help of Eq. (\ref{re-arrangement}) and symmetry properties as the third and fourth terms. Only the prefactor differs by the sign, leading to their cancellation. The part proportional to $\Delta_{ab}$ is canceled in the same way, the major difference lies in the fact that they will be cancelled by their counterpart in $F \leftrightarrow C$. The same applies to the reducible part in order to cancel the completion term $\delta^{\text{S(VP)E}}_{\text{completion}}$.   
\end{widetext}
%
%---------------------------------
%
\subsection{VPVP subset}
The VPVP graph provides only one three-electron contribution, 
\begin{equation}
\Delta E^{(2\text{I})\text{VPVP,3e}}_{v} = \sum_{a,b,i}^{i \neq v} \frac{ I_{v a  i a}(0)  I_{i b v b}(0)}{(\epsilon_v -\epsilon_{i})}\,.
\label{eq:VPVP,3e}
\end{equation}
As has been shown in the one-photon exchange section, it is straightforward to see that the difference between Feynman and Coulomb gauges vanishes since the argument of the photon propagator is zero.
%
%---------------------------------
%
\subsection{V(VP)P subset}
In the case of the V(VP)P graph, there is also only a three-electron contribution which can be found,
\begin{equation}
\Delta E^{(2\text{I})\text{V(VP)P,3e}}_{v} = \sum_{a,b,i}^{i \neq a}  \frac{ I_{v a v i}(0)  I_{i b a b}(0) +I_{v i v a}(0)  I_{a b i b}(0)    }{(\epsilon_a -\epsilon_{i} )}\,.
\label{eq:V(VP)P,3e}
\end{equation}
Similar as in the previous subset, the difference between Feynman and Coulomb gauges is zero due to the zero argument of the photon propagator.
%
%---------------------------------
%
\subsection{V(SE)P subset}
The V(SE)P subset contains the three-electron irreducible contribution $\Delta E^{(2\text{I})\text{V(SE)P,3e,irr}}_{v}$,
\begin{eqnarray}
\Delta E^{(2\text{I})\text{V(SE)P,3e,irr}}_{v} &=& - \sum_{a,b,i}^{i \neq b}\left[ \frac{ I_{v b v i}(0)  I_{a i b a}(\Delta_{ba})}{(\epsilon_b  -\epsilon_{i})}
\right.\nonumber\\&+&\left.   \frac{ I_{v i v b}(0)  I_{a b i a}(\Delta_{ba})}{(\epsilon_b  -\epsilon_{i})}\right]\,,
\label{eq:V(SE)P,3e,irr}
\end{eqnarray}
together with the three-electron reducible term $\Delta E^{(2\text{I})\text{V(SE)P,3e,red}}_{v}$,
\begin{equation}
\Delta E^{(2\text{I})\text{V(SE)P,3e,red}}_{v} = -\sum_{a, b, a_1} I_{v a v a_1}(0)  I^{\prime}_{b a_1 a b}(\Delta_{ab})\,.   
\label{eq:V(SE)P,3e,red}
\end{equation}
As the reducible part arises from the excluded term in the sum, completing the sum will automatically cancel the reducible part. The only subtlety is to break the reducible contribution into two pieces and use the symmetry properties to rewrite it in proper shape.
%
%---------------------------------
%
\subsection{SECP subset}
Let us turn now to the counterpotential contributions and its previously proposed decomposition. Since, after extraction of the interelectronic-interaction parts, all diagrams will be either two-electron (SECP and VPCP subsets) or one-electron (CPCP subset), we do not identify this superscript directly. In the case of the SECP subset, it is easy to obtain the following expression for the corresponding irreducible term $\Delta E^{(2\text{I})\text{SECP,irr}}_{v}$,
\begin{eqnarray}
\Delta E^{(2\text{I})\text{SECP,irr}}_{v} &=& \sum_{a,i}^{i \neq v} \frac{U_{v i} I_{i a a v}(\Delta_{va}) + I_{v a a i}(\Delta_{va}) U_{iv} }{(\epsilon_v -\epsilon_i)}\nonumber\\\nonumber &+& \sum_{a,i}^{i \neq a}   \frac{ I_{v a i v}(\Delta_{va}) U_{ia} + I_{v i a v}(\Delta_{va}) U_{ai} }{(\epsilon_a -\epsilon_i)}\,. \\
\label{eq:SECP,irr}
\end{eqnarray}
The reducible part, which encapsulates also the disconnected SECP contribution of the second term of Eq. (\ref{2nd O cor}), can be written as
\begin{eqnarray}
\Delta E^{(2\text{I}) \text{SECP,red }}_{v} &=&  \sum_{a, v_1} I^{\prime}_{v a a v_1}(\Delta_{va}) U_{v_1 v} \nonumber\\&-& \sum_{a, a_1} I^{\prime}_{v a a_1 v}(\Delta_{va}) U_{a_1 a}\,.
\label{eq:SECP,red}
\end{eqnarray}
It is left to demonstrate the gauge invariance of this subset. As previously, use Eq. (\ref{propagator_difference}), cancel the denominators, and complete the sum one gets for the gauge difference
\begin{eqnarray}
&&\delta\Delta E^{(2\text{I})\text{SECP,irr }}_{v}  = 2\sum_{a,i} \left[ \Delta_{va} \tilde{I}_{v a a i}(\Delta_{va}) U_{i v} \right.\nonumber\\
&&+\left.\Delta_{av} \tilde{I}_{v i a v}(\Delta_{va}) U_{ai}\right] - 2 \sum_{a, v_1} \Delta_{va} \tilde{I}_{v a a v_1}(\Delta_{va}) U_{v_1 v} \nonumber\\
&&- 2 \sum_{a, a_1} \Delta_{av} \tilde{I}_{v a_1 a v}(\Delta_{va}) U_{a a_1}\,. 
\end{eqnarray}
Then Eq. (\ref{prime difference}) applied on the reducible part will provide the appropriate terms to cancel the second line of the previous equation. With $[I(\boldsymbol{x} - \boldsymbol{y}), V(\boldsymbol{z})] = 0$ in mind, a similar formula as Eq. (\ref{re-arrangement}) shows that the complete sum vanishes; therefore, this ends the proof of the gauge invariance.
%
%---------------------------------
%
\subsection{VPCP subset}
The diagrams of the VPCP subset give only one contribution, which yields
\begin{eqnarray}
\Delta E^{(2\text{I})\text{VPCP}}_{v} &=& - \sum_{a,i}^{i \neq v} \frac{ I_{v a i a}(0) U_{i v} + U_{v i} I_{i a v a}(0) }{(\epsilon_v - \epsilon_i)} \nonumber\\ &-& \sum_{a,i}^{i \neq a} \frac{ I_{v a v i}(0) U_{i a} + U_{a i} I_{v i v a}(0) }{(\epsilon_a - \epsilon_i)}\,.
\label{eq:VPCP}
\end{eqnarray}
As the argument of the photon propagator is zero and the counterpotential is not influenced by any gauge change, the gauge invariance of this term is straightforwardly guaranteed. 
%
%---------------------------------
%
\subsection{CPCP subset}
The CPCP subset contains only one term, which corresponds to the diagram itself,
\begin{eqnarray}
\Delta E^{(2\text{I})\text{CPCP}}_{v } = \sum_i^{i \neq v} \frac{U_{v i} U_{i v}}{(\epsilon_v - \epsilon_i)}\,,
\label{eq:CPCP}
\end{eqnarray}
since as in the case of the one-photon exchange, there is no such term in the radiative correction set. By the construction it is gauge invariant.

%=================================================================
%
\section{Results}
\subsection{Final expressions}
\begin{figure}
\centering
\includegraphics[width=0.47\textwidth]{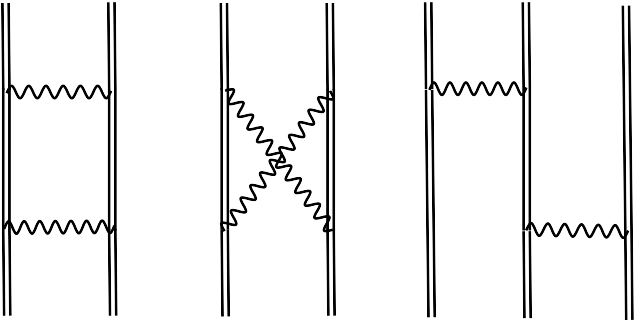}
\caption{The two-photon exchange Feynman diagrams referred to as: the ladder-diagram, the crossed-diagram, and the three-electron diagram. Notations are the same as in Fig. \ref{fig:1}.}
\label{fig:4}
\end{figure}
Thus, in the previous section, we identified eight gauge-invariant subsets in the original Furry picture: SESE (two- and three-electron subsets), SEVP, S(VP)E (two- and three-electron subsets), VPVP, V(VP)P, and V(SE)P. In an extended Furry picture, they are completed by three additional counterpotential subsets: SECP, VPCP, and CPCP. The final expression for the total two-photon exchange correction $\Delta E^{(2\text{I})}_v$,
\begin{widetext}
\begin{eqnarray}
&&\Delta E^{(2\text{I})}_v =
 \left(\Delta E^{(2\text{I})\text{SESE,2e,irr}}_v
      +\Delta E^{(2\text{I})\text{SESE,2e,red}}_v\right)
+\left(\Delta E^{(2\text{I})\text{SESE,3e,irr}}_v
      +\Delta E^{(2\text{I})\text{SESE,3e,red}}_v\right)
+\left(\Delta E^{(2\text{I})\text{SEVP,3e,irr}}_v\right.
\nonumber\\
&&
\left.+\Delta E^{(2\text{I})\text{SEVP,3e,red}}_v\right)
+\left(\Delta E^{(2\text{I})\text{S(VP)E,2e,irr}}_v
      +\Delta E^{(2\text{I})\text{S(VP)E,2e,red}}_v\right)
+\left(\Delta E^{(2\text{I})\text{S(VP)E,3e,irr}}_v
      +\Delta E^{(2\text{I})\text{S(VP)E,3e,red}}_v\right)
\nonumber\\
&&
      +\Delta E^{(2\text{I})\text{VPVP,3e}}_v
      +\Delta E^{(2\text{I})\text{V(VP)P,3e,irr}}_v
+\left(\Delta E^{(2\text{I})\text{V(SE)P,3e,irr}}_v
      +\Delta E^{(2\text{I})\text{V(SE)P,3e,red}}_v\right)
+\left(\Delta E^{(2\text{I})\text{SECP,irr}}_v
      +\Delta E^{(2\text{I})\text{SECP,red}}_v\right)
\nonumber\\
&&
      +\Delta E^{(2\text{I})\text{VPCP}}_v
      +\Delta E^{(2\text{I})\text{CPCP}}_v\,,
\label{eq:total}
\end{eqnarray}
is given by a sum of equations (\ref{eq:SESE,2e,irr}), (\ref{eq:SESE,2e,red}),
(\ref{eq:SESE,3e,irr}), (\ref{eq:SESE,3e,red}), (\ref{eq:SEVP,3e,irr}), (\ref{eq:SEVP,3e,red}), (\ref{eq:S(VP)E,2e,irr}), (\ref{eq:S(VP)E,2e,red}), (\ref{eq:S(VP)E,3e,irr}), (\ref{eq:S(VP)E,3e,red}), (\ref{eq:VPVP,3e}), (\ref{eq:V(VP)P,3e}), (\ref{eq:V(SE)P,3e,irr}), (\ref{eq:V(SE)P,3e,red}), (\ref{eq:SECP,irr}), (\ref{eq:SECP,red}), (\ref{eq:VPCP}), and (\ref{eq:CPCP}). We group the terms in Eq. (\ref{eq:total}) into the gauge-invariance subsets. 

The two-photon exchange corrections can be also identified by the many-electron diagrams depicted in Fig. \ref{fig:4}. All of the contributions originated from these diagrams we obtain within the redefined vacuum approach and present with Eq. (\ref{eq:total}). To compare our results based on the effective one-electron approach with the ones obtained within consideration of many-electron diagrams, we resemble them in accordance with the separation published previously \cite{yerokhin:2001:032109, artemyev:2003:062506, malyshev:2017:022512}. First, we present the expression for the two-electron irreducible contribution $\Delta E^{(2\text{I})\text{2e,irr}}_{v}$, which is obtained by summing up Eqs. (\ref{eq:SESE,2e,irr}) and (\ref{eq:S(VP)E,2e,irr}),
\begin{eqnarray}
\Delta E^{(2\text{I})\text{2e,irr}}_{v} &=& \frac{i}{2\pi} \int d\omega \left\{ \sum_{i,j}^{(i,j) \neq (a,v)}
\frac{I_{v a i j}(\omega) I_{i j v a }(\omega)}{(\epsilon_v - \omega - \epsilon_{i}u)  (\epsilon_a + \omega - \epsilon_{j}u)}
+\sum_{i,j}^\prime
\frac{I_{v j i a}(\omega) I_{i a v j}(\omega)}{(\epsilon_v - \omega - \epsilon_{i}u)(\epsilon_a - \omega - \epsilon_{j}u)}
\right.\nonumber\\
&-&\left.
\sum_{i,j}^{(i,j) \neq (a,v)}
\frac{I_{v a i j}(\omega) I_{i j a v}(\Delta_{va} - \omega)}{(\epsilon_v - \omega - \epsilon_{i}u)(\epsilon_a + \omega - \epsilon_{j}u)}
 - \sum_{i,j}^\prime
\frac{I_{v j i v}(\omega) I_{i a a j}(\Delta_{va} - \omega)}{(\epsilon_v - \omega - \epsilon_{i}u)(\epsilon_v - \omega - \epsilon_{j}u)}\right\}\,. 
\label{eq:2eirred}
\end{eqnarray}
The first two terms are direct contributions [Eq. (\ref{eq:S(VP)E,2e,irr})], ladder and crossed, respectively, while the second two are the exchange counterparts [Eq. (\ref{eq:SESE,2e,irr})], again ladder and crossed, respectively. The primes on the sums are meant for the omitted terms, namely, $i = v\,\&\,j = a$ for the crossed direct term and $(i, j) = \{(a, a), (v, v)\}$ for the crossed exchange term. The two-electron reducible contribution $\Delta E^{(2\text{I})\text{2e,red}}_{v}$ is given by a sum of Eqs. (\ref{eq:SESE,2e,red}) and (\ref{eq:S(VP)E,2e,red}),   
\begin{eqnarray}
\Delta E^{(2\text{I})\text{2e,red}}_{v} &=& \frac{i}{2\pi}\int\frac{d\omega}{(\omega + i \varepsilon)^2}
\left\{ \sum_{a, a_1, v_1} \left[
   I_{v a v_1 a_1}(\omega) I_{v_1 a_1 a v}(\Delta_{va}+\omega)
 + I_{v a a_1 v_1}(\Delta_{va}-\omega) I_{a_1 v_1 a v}(\omega)
\right.\right.\nonumber\\
&-&\left.
   I_{v a a_1 v_1 }(\Delta_{va}-\omega) I_{a_1 v_1 v a}(\Delta_{va}-\omega)/2
 - I_{v a a_1 v_1 }(\Delta_{va}+\omega) I_{a_1 v_1 v a}(\Delta_{va}+\omega)/2\right]
\nonumber\\
&-&\left.
  \sum_{a, a_1, a_2} I_{v a_2 a_1 v}(\Delta_{va}- \omega) I_{a_1 a a a_2} (\omega)
- \sum_{a, v_1, v_2} I_{v v_2 v_1 v}(\omega) I_{v_1 a a v_2}(\omega + \Delta_{va})
\right\}\,.
\label{eq:2ered}
\end{eqnarray}
The expressions of the two-electron contributions presented in such a way agree with the results of Refs. \cite{yerokhin:2001:032109, artemyev:2003:062506, malyshev:2017:022512}. In Refs. \cite{sapirstein:2011:012504, sapirstein:2015:062508}, the two-electron contributions are not separated into irreducible and reducible parts and given as our Eq. (\ref{eq:2eirred}) but without any sums' restrictions. Let us recall the separations made between irreducible and reducible terms. Concerning the two-electron crossed terms, the reducible contributions are separated artificially and can be added back, thus, removing the restrictions in the second and fourth sums in Eq. (\ref{eq:2eirred}). In contrast, the two-electron ladder term is leading to a discrepancy. In order to restore the completeness of the first and third sums in Eq. (\ref{eq:2eirred}), the omitted contributions $(i,j) = (a,v)$ should have the pole structure $(\omega + i\varepsilon)(\omega - i\varepsilon)$. However, as one can see from Eq. (\ref{eq:2ered}) all contributions have a different pole topology, namely, $(\omega + i\varepsilon)^2$.

Collecting all three-electron irreducible terms given by Eqs. (\ref{eq:SESE,3e,irr}), (\ref{eq:SEVP,3e,irr}), (\ref{eq:S(VP)E,3e,irr}), (\ref{eq:VPVP,3e}), (\ref{eq:V(VP)P,3e}), and (\ref{eq:V(SE)P,3e,irr}) together with the counterpotential irreducible terms, Eqs. (\ref{eq:SECP,irr}), (\ref{eq:VPCP}), and (\ref{eq:CPCP}), we come to the following total three-electron contribution
\begin{eqnarray}
\Delta E^{(2\text{I})\text{3e,irr}}_{v} &=&
\sum_{a,b,i}^{i \neq a} \frac{\left[I_{i b a b}(0) - I_{i b b a}(\Delta_{ab})\right] \left[I_{v a v i }(0) - I_{v a i v}(\Delta_{va})\right] + \left[I_{a b i b}(0) - I_{b a i b}(\Delta_{ab})\right] \left[I_{v i v a}(0) - I_{v i a v}(\Delta_{va})\right]    }{\epsilon_a - \epsilon_i}\nonumber\\
&+&
\sum_{a,b,i}^{i \neq v} \frac{\left[I_{v a a i}(\Delta_{va}) - I_{v a i a}(0)\right] \left[I_{i b b v}(\Delta_{vb}) - I_{i b v b}(0)\right]}{\epsilon_v - \epsilon_i}
+\sum_{a,b,i} \frac{\left[I_{a b i v}(\Delta_{vb}) - I_{a b v i}(\Delta_{va})\right] I_{v i a b}(\Delta_{va})}{\epsilon_a + \epsilon_b - \epsilon_v - \epsilon_i}\nonumber\\ 
&+&
\sum_{a,b,i}^{(i,b) \neq (v,a)} \frac{\left[I_{v a b i}(\Delta_{vb}) - I_{v a i b}(\Delta_{ba})\right] \left[I_{b i a v}(\Delta_{ba}) - I_{b i v a}(\Delta_{vb})\right]}{\epsilon_v + \epsilon_a - \epsilon_b - \epsilon_i}\nonumber\\
&+&
\sum_{a,i}^{i \neq v} \frac{U_{vi} \left[I_{i a a v }(\Delta_{va}) - I_{i a v a}(0) \right] + \left[I_{v a a i}(\Delta_{va}) - I_{v a i a}(0)\right] U_{i v}}{\epsilon_v - \epsilon_i}\nonumber\\
&+&
\sum_{a,i}^{i \neq a} \frac{U_{a i} \left[I_{v i a v}(\Delta_{va}) - I_{v i v a}(0)\right] + \left[I_{v a i v}(\Delta_{va}) - I_{v a v i}(0)\right] U_{i a}}{\epsilon_a - \epsilon_i} 
+\sum_i^{i \neq v} \frac{U_{v i} U_{i v}}{\epsilon_v - \epsilon_i}\,.
\label{eq:3e,irr}
\end{eqnarray}
In the same way, summing Eqs. (\ref{eq:SESE,3e,red}), (\ref{eq:SEVP,3e,red}), (\ref{eq:S(VP)E,3e,red}), (\ref{eq:V(SE)P,3e,red}), and (\ref{eq:SECP,red}) one derives
the total three-electron reducible contribution:
\begin{eqnarray}
\Delta E^{(2\text{I})\text{3e,red}}_{v} &=&
\sum_{a, b, a_1}\left\{
I^{\prime}_{v a_1 a v}(\Delta_{va})\left[I_{a b a_1 b}(0) - I_{ a b b a_1}(\Delta_{ab})
\right] + I^{\prime}_{a b b a_1}(\Delta_{ab})\left[I_{v a_1 a v}(\Delta_{va}) - I_{v a_1 v a}(0)\right]\right\}
\nonumber\\
&+&\sum_{a, b, v_1}
I^{\prime}_{v a a v_1}(\Delta_{va})\left[I_{v_1 b b v}(\Delta_{vb}) - I_{v_1 b v b}(0)\right]
- \sum_{a, a_1} I^{\prime}_{v a a_1 v}(\Delta_{va}) U_{a_1 a}
+ \sum_{a, v_1} I^{\prime}_{v a a v_1}(\Delta_{va}) U_{v_1 v}
\nonumber\\
&+&\sum_{a, a_1, v_1}
I^{\prime}_{v a a_1 v_1}(\Delta_{va})\left[I_{a_1 v_1 a v}(0) - I_{a_1 v_1 v a}(\Delta_{va})\right]\,.
\label{eq:3e,red}
\end{eqnarray}
\end{widetext} 
In contrast to the two-electron contributions, the three-electron terms have a more complicated structure and can not be easily generalized. Therefore, they were mainly derived for a particular state, e.g., for Li-like ions in Refs. \cite{yerokhin:2001:032109, artemyev:2003:062506} or B-like ions in Ref. \cite{malyshev:2017:022512}, and for these special cases our expressions agree with the mentioned papers. However, in Refs. \cite{sapirstein:2011:012504, sapirstein:2015:062508}, the general expressions for the three-electron terms are given, e.g., $E_{2F}$ as the irreducible part and $E_{2F'}$ as the reducible part in accordance with Ref. \cite{sapirstein:2015:062508}. Comparing our expressions (\ref{eq:3e,irr}) and (\ref{eq:3e,red}) with those of Sapirstein and Cheng, we find an almost perfect agreement for the irreducible contribution, except for an absence of a restriction in the fourth sum in Eq. (\ref{eq:3e,irr}). Whereas the reducible contribution strongly disagrees. Comparing term by term, we notice problems with contributions in the first and third lines of Eq. (\ref{eq:3e,red}). For example, the last term is written with $b$ state instead of $a_1$, meaning the summations over all core electrons independently. The first term in the first line is written with apparent misprints (odd number of electron states in matrix elements). Moreover, there is no contribution like our second term in the first line in Ref. \cite{sapirstein:2015:062508}. Here, we should note that the gauge invariance condition dictates the exact form of the reducible contributions. The detailed analysis of gauge invariance for each subset of diagrams performed here makes us believe that our result is more reliable.
\subsection{Numerical values}
Now we are in a position to support numerically the gauge invariance of the identified subsets -- the strength of the developed approach. The computations are performed for the $2s$, $2p_{1/2}$, and $2p_{3/2}$ binding energies in Li-like tin ion. The finite nuclear size is accounted for by using the Fermi model for the nucleus charge distribution with the root-mean-square radius of 4.6519 fm \cite{angeli:2013:69}. The $\omega$ integration in the two-electron terms was performed with the Wick rotation to the integration contours chosen as in Ref. \cite{mohr:2000:052501}. The infinite summations over the whole Dirac spectrum are performed using the dual-kinetic-balance finite basis set method \cite{shabaev:2004:130405} for the Dirac equation. The basis functions are constructed from B-splines \cite{sapirstein:1996:5213}. The number of basis functions is systematically increased to achieve a clear convergence pattern of the calculated results. Then the extrapolation to an infinite number of basis functions is undertaken. The partial wave summation over the Dirac quantum number $\kappa$ was terminated at $\kappa_{\rm max} = 8$, and the remainder was estimated by the least-squares inverse polynomial fitting. The numerical uncertainty is estimated to be less than $3 \times 10^{-6}$ atomic units.

The numerical evaluations are performed in both Feynman and Coulomb gauges to demonstrate the gauge invariance of the identified subsets. The individual contributions and sums for each subset are presented in Tables \ref{tab:CO} and \ref{tab:CH} in atomic units. In Table \ref{tab:CO}, the results for the case of Coulomb potential (original Furry picture) are presented. As one can see from the table, the values for each of eight identified subsets are gauge invariant within the numerical uncertainty. We notice that the term $\Delta E^{(2\text{I})\text{V(SE)P,3el,red}}$ is not displayed, since it gives zero contribution in the case of Li-like ions, i.e., ions with only one closed shell. For a comparison, we also present results from Ref. \cite{yerokhin:2001:032109} for the $2s$ and $2p_{1/2}$ states and from \cite{artemyev:2003:062506} for the $2p_{3/2}$ state. The comparison shows an excellent agreement with values obtained in the present work. In Table \ref{tab:CH}, we also display the results for the extended Furry picture. The self-consistent core-Hartree potential has been used in this case. In addition to the subsets presented already in the original Furry picture case, we have three additional subsets coming from the counterpotential diagrams. Again, as in the former case, the values presented in this table confirm all identified subsets' gauge invariance.

Presenting now the accurate control on numerical results, we perform calculations for the most compelling experimental cases; Li-like bismuth and uranium ions are presented. In Table \ref{tab:comparison}, the ionization potentials and transition energies for the $2s$ and $2p_{3/2}$ states in the case of Li-like Bi and for the $2s$ and $2p_{1/2}$ states in the case of Li-like U ions. The following nuclear radii $5.531$ fm and $5.860$ fm are employed for bismuth and uranium, respectively, following Ref. \cite{sapirstein:2011:012504}. As in Ref. \cite{sapirstein:2011:012504} we use the Kohn-Sham potential as the starting potential and compare term by term the zeroth-order energy $E^{0}$ as well as the first-order $\Delta E^{(1)}$ and second-order $\Delta E^{(2)}$ corrections. As one can see from the table, the zeroth- and first-order terms are in perfect agreement to all given digits. At the same time, the two-photon exchange (second-order) correction is slightly different. The reason for this may be precisely the discrepancy in the reducible two- and three-electron contributions stated above. Since the reducible contributions are always much smaller than the dominant irreducible one it is rather difficult to identify the correctness of its calculation. Here, we should emphasize again that the reducible contributions are responsible for restoring the gauge invariance, and within the developed method, we have excellent control on this issue.

To compare with the experiments, we have to add additional correction $\Delta E_{\rm rest}$, which incorporates three-photon exchange, recoil, radiative, and screened radiative contributions. These contributions are taken directly from Ref. \cite{sapirstein:2011:012504}. Comparing now the total values of the energies for the $2p_{3/2} - 2s$ transition in bismuth and the $2p_{1/2} - 2s$ transition in uranium, one can see overall good agreement between both theoretical calculations and experimental results within the given error bars. Nevertheless, we believe that our independent calculations are essential for future more precise comparisons.
%
%=================================================================
%
\section{Conclusion}
The redefined vacuum approach within the bound-state QED was presented in detail and applied to the derivation of the one- and two-photon exchange corrections for atoms with one valence electron over closed-shells. The general formulas for the two-photon exchange correction are presented. The employment of the redefined vacuum approach allowed us to identify the gauge-invariant subsets at two- and three-electron diagrams and separate between the direct and exchange contributions at two-electron graphs. In total, we identify eight gauge-invariant subsets in the case of the original Furry picture. In the case of the screening potential involved, an additional three subsets are originated from the counterpotential diagrams. The gauge invariance of the identified subsets is verified analytically for three-electron terms and numerically for all of the subsets on the example of Li-like ions. The possibility of checking the gauge invariance allows us to control the correctness of the derived expressions and verify the numerical calculations by comparing the results for each identified subset in different gauges. The presented redefined vacuum approach can be further employed for atoms with a more complicated electronic structure. Moreover, the identification of gauge-invariant contributions within this approach paves the way for calculating the higher-order corrections, which can be split into gauge-invariant subsets and tackled one after the other.
%
%=================================================================
%
\section{Acknowledgments}
The communications with Jonathan Sapirstein are appreciated. This work was supported by DFG (VO1707/1-3) and by RFBR (Grant No. 19-02-00974). A.V.V. acknowledges financial support by the Government of the Russian Federation through the ITMO Fellowship and Professorship Program. D.A.G. acknowledges the support by the Foundation for the Advancement of Theoretical Physics and Mathematics ``BASIS.''
%
%=================================================================
%
\begin{table*}
\caption{Contributions of the identified gauge-invariant subsets to the two-photon exchange corrections for the $2s$, $2p_{1/2}$, and $2p_{3/2}$ states of the Li-like Sn ($Z = 50$) ion, in atomic units. The values are obtained within the original Furry picture (Coulomb potential). The results of the Feynman and Coulomb gauges are given. The Fermi model of the nuclear charge distribution is employed.}
\label{tab:CO}
\begin{center}
\tabcolsep10pt
\begin{tabular}{l l l l l l l}
\hline\hline 
 &  \multicolumn{2}{c}{$2s$} &  \multicolumn{2}{c}{$2p_{1/2}$} &  \multicolumn{2}{c}{$2p_{3/2}$} \\  [1ex]
 & Feynman & Coulomb & Feynman & Coulomb & Feynman & Coulomb\\ [1ex]
\hline      \\[0.5ex]            
SESE,2e,irr & 0.026650 &  \ 0.029926 &  0.022374 & \ 0.025697 & 0.027738 & 0.031426 \\[3ex] 
SESE,2e,red & 0.003036 & -0.000240 & 0.002901  & -0.000422 & 0.003839 & 0.000151   \\[3ex]
SESE,2e & 0.029686 & \ 0.029686 & 0.025275 & \ 0.025275 & 0.031578 &  0.031578 \\[1ex]
 &  0.02968$^{a}$ & \ 0.02967$^{a}$ $^{*}$  & 0.02528$^{a}$ & \ 0.02529$^{a}$ & 0.03156$^{b}$ & \\
\hline      \\[0.5ex] 
S(VP)E,2e,irr & -0.187925 & -0.187589  & -0.323212 & -0.323137 & -0.212025 & -0.211837 \\[3ex] 
S(VP)E,2e,red & \ 0.000338 & \ 0.000001 & \ 0.000052 & -0.000022 & \ 0.000187 & -0.000002 \\[3ex] 
S(VP)E,2e     & -0.187588 & -0.187587  & -0.323159 & -0.323159 & -0.211838 &  -0.211838 \\[1ex]
 & -0.18759$^{a}$ & -0.18759$^{a}$ &   -0.32317$^{a}$ & -0.32317$^{a}$ & -0.21185$^{b}$ & \\
\hline      \\[0.5ex] 
SESE,3e,irr &  -0.026028 & -0.027012 & -0.015884 & -0.017989 & -0.022614 & -0.024791 \\[3ex] 
SESE,3e,red & -0.000933  & \ 0.000052 & -0.002146 & -0.000041 & -0.002602 & -0.000424 \\[3ex]
SESE,3e     &  -0.026960 & -0.026960 & -0.018030 & -0.018030 & -0.025215 &  -0.025215\\[1ex]
\hline      \\[0.5ex] 
SEVP,3e,irr & 0.035264  & \ 0.037506 & 0.018094 & 0.022872 & 0.037953& 0.042854 \\[3ex] 
SEVP,3e,red & 0.002122  & -0.000120 & 0.004866 & 0.000088 & 0.005850 &  0.000949 \\[3ex]  
SEVP,3e     &  0.037386 & \ 0.037386 & 0.022960 & 0.022960 & 0.043803 & 0.043803 \\[1ex]
\hline      \\[0.5ex] 
S(VP)E,3e,irr  & \ 0.176478  &  0.176348 & \ 0.305829 & 0.305787 & \ 0.197613 & \ 0.197495 \\[3ex] 
S(VP)E,3e,red  &  -0.000130 & 0.000000 & -0.000041 & 0.000001 & -0.000140 & -0.000022 \\[3ex] 
S(VP)E,3e      & \ 0.176348  &  0.176348 & \ 0.305788 & 0.305788  & \ 0.197473 & \ 0.197473 \\[1ex]
\hline      \\[0.5ex] 
VPVP,3e    & -0.302978  & -0.302978 & -0.483510 & -0.483510 & -0.430804 & -0.430804 \\[1ex]  
\hline      \\[0.5ex] 
V(VP)P,3e & -0.050230  & -0.050230 & -0.038836 & -0.038836 & -0.029621 & -0.029621  \\[1ex] 
\hline      \\[0.5ex] 
V(SE)P,3e,irr  &  0.021757 & 0.021757 & 0.017132 & 0.017132 & 0.013100 & 0.013100 \\ [1ex]
\hline      \\[0.5ex] 
Total 3-electron & -0.144678  & -0.144678 & -0.194497 & -0.194497  & -0.231265 & -0.231265 \\[1ex]
 & -0.14468$^{a}$ & -0.14468$^{a}$ & -0.19450$^{a}$ & -0.19449$^{a}$ & -0.23126$^{b}$ & \\ 
 \\ 
Total & -0.302579 & -0.302579 & -0.492381 & -0.492381 & -0.411525 & -0.411526 \\[1ex]
&-0.30259$^{a}$ & -0.30259$^{a}$ & -0.49239$^{a}$ & -0.49237$^{a}$ & -0.41156$^{b}$ & \\
 \hline \hline
\multicolumn{7}{l}{\footnotesize{$^{a}$ Yerokhin {\it et al.} \cite{yerokhin:2001:032109}; sphere model for the nuclear charge distribution used.}} \\
\multicolumn{7}{l}{\footnotesize{$^{b}$ Artemyev {\it et al.} \cite{artemyev:2003:062506}; sphere model for the nuclear charge distribution used.}} \\
\multicolumn{7}{l}{\footnotesize{$^{*}$ Misprint corrected: Table III, $\Delta E_{\text{ir}}^{\text{2el}} \text{(exch)}$ 0.02995 a.u. is used instead of 0.02595 a.u.}} \\
\end{tabular}
\end{center}
\end{table*}
%
\begin{comment}
\begin{table*}
\caption{Individual two-electron contributions to the two-photon exchange corrections for the $2s, 2p_{1/2}$ and $2p_{3/2}$ state of Li-like $\ce{^{120}_{50}Sn}$ in extended Furry picture with core Hartree counterpotential in the Feynman and Coulomb gauge using Fermi model of the nuclear charge distribution, in atomic units.  }
\begin{center}
\begin{tabular}{c c c c c c c } 
\hline\hline 
 &  \multicolumn{2}{c}{2s} &  \multicolumn{2}{c}{2p_{1/2}} &  \multicolumn{2}{c}{2p_{3/2}} \\  [1ex]
 & Feynman & Coulomb & Feynman & Coulomb & Feynman & Coulomb\\ [1ex]
\hline      \\[0.5ex]            
\Delta E^{(2\text{I}) \text{SESE,2el,irr}}_{v} & 0.0271088 & 0.0302927  &  0.0230037
 & 0.0261824 & 0.027727 & 0.0312271 \\[3ex] 
\Delta E^{(2\text{I})\text{SESE,2el,red}}_{v} & 0.0029577 & -0.00022637 & 0.00278486
  & -0.000393687 & 0.00363065 &  0.0001301  \\[3ex]
\Delta E^{(2\text{I})\text{SESE,2el}}_{v} & 0.0300665 &  0.0300663 & 0.0257886 & 0.0257887 &0.0313576 &0.0313572  \\[1ex]
\hline      \\[0.5ex] 
\Delta E^{(2\text{I})\text{S(VP)E,2el,irr}}_{v} & -0.189003 &  -0.188689 & -0.3247 & -0.324633 & -0.214022 & -0.213855 \\[3ex] 
\Delta E^{(2\text{I})\text{S(VP)E,2el,red}}_{v} & 0.000316434 & 0.00000119  & 0.0000476 & -0.000019494 & 0.000164402 & -0.000001816 \\[3ex] 
\Delta E^{(2\text{I}) \text{S(VP)E,2el}}_{v}  & -0.188687 &  -0.188688 & -0.324652 & -0.324652 & -0.213857 & -0.213857  \\[1ex]
\hline \hline
\end{tabular}
\label{table:nonlin}
\end{center}
\end{table*}
\end{comment}
%
\begin{table*}
\caption{Contributions of the identified gauge-invariant subsets to the two-photon exchange corrections for the $2s$, $2p_{1/2}$, and $2p_{3/2}$ states of the Li-like Sn ($Z = 50$) ion, in atomic units. The values are obtained within the extended Furry picture (core-Hartree potential). The results of the Feynman and Coulomb gauges are given. The Fermi model of the nuclear charge distribution is employed.}
\label{tab:CH}
\begin{center}
\tabcolsep10pt
\begin{tabular}{l l l l l l l}
\hline\hline 
 &  \multicolumn{2}{c}{$2s$} &  \multicolumn{2}{c}{$2p_{1/2}$} &  \multicolumn{2}{c}{$2p_{3/2}$} \\  [1ex]
 & Feynman & Coulomb & Feynman & Coulomb & Feynman & Coulomb\\ [1ex]
\hline      \\[0.5ex] 
SESE,2e,irr & 0.027109 & \ 0.030293  &  0.023004 & \ 0.026182 & 0.027727 & 0.031227 \\[3ex] 
SESE,2e,red & 0.002958 & -0.0002264 & 0.002785  & -0.000394 & 0.003631 &  0.000130  \\[3ex]
SESE,2e     & 0.030067 & \ 0.030066 & 0.025789 & \ 0.025789 &0.031358 &0.031357  \\[1ex]
\hline      \\[0.5ex] 
S(VP)E,2e,irr & -0.189003 &  -0.188689 & -0.324700 & -0.324633 & -0.214022 & -0.213855 \\[3ex] 
S(VP)E,2e,red & \ 0.000316 & \ 0.000001  & \ 0.000048 & -0.000019 & \ 0.000164 & -0.000002 \\[3ex] 
S(VP)E,2e     & -0.188687 &  -0.188688 & -0.324652 & -0.324652 & -0.213857 & -0.213857  \\[1ex]
\hline 
SESE,3e,irr  & -0.026394 & -0.027326 & -0.016618 & -0.018611 & -0.022634 & -0.024685 \\[3ex] 
SESE,3e,red  & -0.000886  & \ 0.000048 & -0.002033 & -0.000040 & -0.002437 & -0.000386 \\[3ex]
SESE,3e      &  -0.027280 & -0.027280 & -0.018651 & -0.018651 & -0.025071 & -0.025071 \\[1ex]
\hline      \\[0.5ex] 
SEVP,3e,irr  &  0.035830 & \ 0.037949 & 0.019778 & 0.024279 & 0.038359 & 0.042952 \\[3ex] 
SEVP,3e,red  &  0.002009 & -0.000110 & 0.004585 & 0.000085 & 0.005452 & 0.000859  \\[3ex]  
SEVP,3e      & 0.037839  & \ 0.037839 & 0.024364 & 0.024364 & 0.043811  & 0.043811 \\[1ex]
\hline      \\[0.5ex] 
S(VP)E,3e,irr  & \ 0.177970 &  0.177851 & \ 0.308182 & 0.308145 & \ 0.200231 &  \ 0.200128 \\[3ex] 
S(VP)E,3e,red  & -0.000119 & 0.000000 & -0.000037  & 0.000001 & -0.000122 & -0.000018 \\[3ex] 
S(VP)E,3e     & \ 0.177851 &  0.177851 & \ 0.308145 &  0.308146 & \ 0.200109 & \ 0.200110 \\[1ex]
\hline      \\[0.5ex] 
VPVP,3e  &  -0.306234 & -0.306234 & -0.488841 & -0.488841 & -0.436894 & -0.436894 \\[1ex]  
\hline      \\[0.5ex] 
V(VP)P,3e &  -0.049365 & -0.049365 & -0.036796 & -0.036796 & -0.028280 & -0.028280  \\[1ex] 
\hline      \\[0.5ex] 
V(SE)P,3e,irr  & 0.021462 & 0.021462 & 0.016290 & 0.016290 & 0.012550 & 0.012550  \\ [1ex]
 \hline   \\[0.5ex] 
SECP,irr  &  -0.035830 & -0.037949 & -0.019778 & -0.024279 & -0.038359 & -0.042952 \\[3ex] 
SECP,red & -0.002009  & \ 0.000110 & -0.004585 & -0.000085 & -0.005452 & -0.000859 \\[3ex] 
SECP     & -0.037839  & -0.037839 & -0.024364 & -0.024364 & -0.043811 & -0.043811  \\[1ex]
\hline      \\[0.5ex] 
VPCP     & 0.661834  & 0.661834 & 1.014480 & 1.014480 & 0.902069 & 0.902069 \\[0.5ex] 
\hline      \\[0.5ex] 
CPCP  &  \multicolumn{2}{c}{-0.306235}   &  \multicolumn{2}{c}{-0.488842}   & \multicolumn{2}{c}{-0.436895}  \\[0.5ex]
%\hline      \\[0.5ex]  &  \multicolumn{2}{c}{2s} &  \multicolumn{2}{c}{2p_{1/2}} &  \multicolumn{2}{c}{2p_{3/2}} \\  [1ex]
%\text{3el}    & 0.172032  & 0.172032 & 0.305784 &  0.305784 & 0.187588 & 0.187588
% \\ [0.5ex]
\hline \\ [0.5ex]
Total & 0.013412 & 0.013410& 0.006921 & 0.006921 & 0.005089 & 0.005088 \\
\hline \hline
\end{tabular}
\end{center}
\end{table*}

\begin{table*}
\caption{Individual contributions to the ionization potentials (a.u.) and transition energies (a.u.) for the $2s$ and $2p_{3/2}$ states in the case of Li-like Bi (upper half) and for the $2s$ and $2p_{1/2}$ states in case of the Li-like U (lower half). The Dirac energy $E^{(0)}$, one-photon $\Delta E^{(1)}$ and two-photon $\Delta E^{(2)}$ exchange corrections are compared with the results of Ref. \cite{sapirstein:2011:012504}. The Kohn-Sham potential is employed as the starting potential. Other correction $\Delta E_{\rm rest}$, which includes three-photon exchange, recoil, radiative, and screened radiative contributions, is taken from Ref. \cite{sapirstein:2011:012504} and also used for calculation of our total theory results. A comparison with existing experimental values is given.}
\label{tab:comparison}
\begin{center}
\tabcolsep10pt
\begin{tabular}{lr@{}lr@{}lr@{}lr@{}lr@{}lr@{}l}
\hline\hline 
$Z = 83$ & \multicolumn{4}{c}{$2s$} & \multicolumn{4}{c}{$2p_{3/2}$} & \multicolumn{4}{c}{$2p_{3/2} - 2s$} \\
& \multicolumn{2}{c}{Ref. \cite{sapirstein:2011:012504}} & \multicolumn{2}{c}{This work}
& \multicolumn{2}{c}{Ref. \cite{sapirstein:2011:012504}} & \multicolumn{2}{c}{This work}
& \multicolumn{2}{c}{Ref. \cite{sapirstein:2011:012504}} & \multicolumn{2}{c}{This work} \\ \hline\\[-2ex]
$E^{(0)}$             & $-$937.&4309     & $-$937.&43089
                      & $-$833.&7473     & $-$833.&74735
                      &    103.&6836     &    103.&68354 \\
$\Delta E^{(1)}$      &   $-$6.&5674     &   $-$6.&56746
                      &   $-$6.&8275     &   $-$6.&82756
                      &   $-$0.&2601     &   $-$0.&26010 \\
$\Delta E^{(2)}$      &   $-$0.&0178     &   $-$0.&01725
                      &   $-$0.&0159     &   $-$0.&01491
                      &      0.&0019     &      0.&00234 \\
$\Delta E_{\text{rest}}$
                      &      1.&1413     &      1.&1413
                      &      0.&1747     &      0.&1747
                      &   $-$0.&9666(18) &   $-$0.&9666(18) \\
$E_{\text{theo}}$     & $-$942.&8748     & $-$942.&87431
                      & $-$840.&4160     & $-$840.&41512
                      &    102.&4588(18) &    102.&4592(18) \\[0.5ex]
$E_{\text{exp}}$ \cite{beiersdorfer:1998:3022}
                      &        &         &        &
                      &        &         &        &
                      &\multicolumn{4}{c}{102.4622(14)} \\ \hline \\[-2ex]
$Z=92$ & \multicolumn{4}{c}{$2s$} & \multicolumn{4}{c}{$2p_{1/2}$} &  \multicolumn{4}{c}{$2p_{1/2} - 2s$ } \\
& \multicolumn{2}{c}{Ref. \cite{sapirstein:2011:012504}} & \multicolumn{2}{c}{This work}
& \multicolumn{2}{c}{Ref. \cite{sapirstein:2011:012504}} & \multicolumn{2}{c}{This work}
& \multicolumn{2}{c}{Ref. \cite{sapirstein:2011:012504}} & \multicolumn{2}{c}{This work} \\ \hline\\[-2ex]
$E^{(0)}$             & $-$1201.&067    & $-$1201.&06660 
                      & $-$1192.&233    & $-$1192.&23254  
                      &       8.&834    &       8.&83406 \\
$\Delta E^{(1)}$      &    $-$7.&360    &    $-$7.&36038 
                      &    $-$4.&305    &    $-$4.&30548  
                      &       3.&055    &       3.&05490 \\
$\Delta E^{(2)}$      &    $-$0.&023    &    $-$0.&02187 
                      &    $-$0.&070    &    $-$0.&06914  
                      &    $-$0.&047    &    $-$0.&04727 \\
$\Delta E_{\text{rest}}$
                      &       1.&731    &       1.&731 
                      &       0.&202    &       0.&202
                      &    $-$1.&529(3) &    $-$1.&529(3) \\ 
$E_{\text{theo}}$     & $-$1206.&719    & $-$1206.&71785 
                      & $-$1196.&406    & $-$1196.&40515 
                      &      10.&313(3) & 10.&3127(30)    \\[0.5ex]
$E_{\text{exp}}$ \cite{beiersdorfer:2005:233003}
                      &         &       &        &
                      &         &       &        &
                      & \multicolumn{4}{c}{10.3135(5)} \\ \hline\hline
\end{tabular}
\end{center}
\end{table*}

\bibliography{bibliography}
%\addcontentsline{toc}{section}{References}
%=================================================================
\end{document}